\documentclass[twocolumn,prb,showpacs,superscriptaddress,amsmath,amssymb,natbib]{revtex4}

\usepackage{graphicx}
\usepackage[english,german]{babel}
\usepackage{amsmath}
\usepackage{amssymb}
\usepackage{subfigure}
\usepackage[latin1]{inputenc}

\begin{document}

\title{A Simulation Study of the Lithium Ion Transport Mechanism in Ternary Polymer Electrolytes -- The Critical Role of the Segmental Mobility}

\author{Diddo Diddens}
\email{d.diddens@uni-muenster.de}
\affiliation{Institut f\"ur physikalische Chemie, Westf\"alische Wilhelms-Universit\"at M\"unster, Corrensstrasse 28/30, 48149 M\"unster, Germany}
\affiliation{NRW Graduate School of Chemistry, Corrensstrasse 36, 48149 M\"unster, Germany}

\author{Andreas Heuer}
\affiliation{Institut f\"ur physikalische Chemie, Westf\"alische Wilhelms-Universit\"at M\"unster, Corrensstrasse 28/30, 48149 M\"unster, Germany}
\affiliation{NRW Graduate School of Chemistry, Corrensstrasse 36, 48149 M\"unster, Germany}

\selectlanguage{english}

\date{\today}

\newcommand{\eg}{\mbox{e.\,g.\,}\ }
\newcommand{\ie}{\mbox{i.\,e.\,}\ }
\newcommand{\etal}{{\it et al.\,}}

\newcommand{\Li}{{$\text{Li}^\text{+}$}}

\newcommand{\PnoIL}{{$\text{PEO}_{20}\text{LiTFSI}$}}
\newcommand{\PsomeIL}{{$\text{PEO}_{20}\text{LiTFSI}\cdot0.66\text{~PYR$_\mathrm{13}$TFSI}$}}
\newcommand{\PmoreIL}{{$\text{PEO}_{20}\text{LiTFSI}\cdot3.24\text{~PYR$_\mathrm{13}$TFSI}$}}
\newcommand{\PxIL}{{$\text{PEO}_{20}\text{LiTFSI}\cdot\,x\text{~PYR$_\mathrm{13}$TFSI}$}}

\newcommand{\Ptwenty}{{$\text{PEO}_{20}\text{LiTFSI}$}}
\newcommand{\Psixteen}{{$\text{PEO}_{16}\text{LiTFSI}\cdot0.556\text{~PYR$_\mathrm{13}$TFSI}$}}
\newcommand{\Ptwelve}{{$\text{PEO}_{12}\text{LiTFSI}\cdot1.111\text{~PYR$_\mathrm{13}$TFSI}$}}
\newcommand{\Peight}{{$\text{PEO}_{8}\text{LiTFSI}\cdot1.667\text{~PYR$_\mathrm{13}$TFSI}$}}
\newcommand{\Pzero}{{$\text{PYR}_\mathrm{13}\text{TFSI}\cdot0.262\text{~LiTFSI}$}}
\newcommand{\PminusxIL}{{$\text{PEO}_{20-\alpha x}\text{LiTFSI}\cdot\,x\text{~PYR$_\mathrm{13}$TFSI}$}}

\newcommand{\Pno}{{$\text{P}_{20}\text{S}$}}
\newcommand{\Psome}{{$\text{P}_{20}\text{S}\cdot0.66\text{~IL}$}}
\newcommand{\Pmore}{{$\text{P}_{20}\text{S}\cdot3.24\text{~IL}$}}
\newcommand{\Px}{{$\text{P}_{20}\text{S}\cdot\,x\text{~IL}$}}

\newcommand{\Ptwen}{{$\text{P}_{20}\text{S}$}}
\newcommand{\Psixt}{{$\text{P}_{16}\text{S}\cdot0.556\text{~IL}$}}
\newcommand{\Ptwel}{{$\text{P}_{12}\text{S}\cdot1.111\text{~IL}$}}
\newcommand{\Pei}{{$\text{P}_{8}\text{S}\cdot1.667\text{~IL}$}}
\newcommand{\Pz}{{$\text{IL}\cdot0.262\text{~S}$}}
\newcommand{\Pminusx}{{$\text{P}_{20-\alpha x}\text{S}\cdot\,x\text{~IL}$}}

\newcommand{\Ptwentyonezero}{{$\text{PEO}_{20}\text{LiTFSI}\cdot0.0\text{~PYR$_\mathrm{14}$TFSI}$}}
\newcommand{\Ptwentyoneone}{{$\text{PEO}_{20}\text{LiTFSI}\cdot1.0\text{~PYR$_\mathrm{14}$TFSI}$}}
\newcommand{\Ptwentyonetwo}{{$\text{PEO}_{20}\text{LiTFSI}\cdot2.0\text{~PYR$_\mathrm{14}$TFSI}$}}
\newcommand{\Ptwentyonefour}{{$\text{PEO}_{20}\text{LiTFSI}\cdot4.0\text{~PYR$_\mathrm{14}$TFSI}$}}
\newcommand{\Ptenonezero}{{$\text{PEO}_{10}\text{LiTFSI}\cdot0.0\text{~PYR$_\mathrm{14}$TFSI}$}}
\newcommand{\Ptenoneone}{{$\text{PEO}_{10}\text{LiTFSI}\cdot1.0\text{~PYR$_\mathrm{14}$TFSI}$}}
\newcommand{\Ptenonetwo}{{$\text{PEO}_{10}\text{LiTFSI}\cdot2.0\text{~PYR$_\mathrm{14}$TFSI}$}}
\newcommand{\Pfiveoneone}{{$\text{PEO}_{5}\text{LiTFSI}\cdot1.0\text{~PYR$_\mathrm{14}$TFSI}$}}

\newcommand{\Ptwonezero}{{$\text{P}_{20}\text{S}$}}
\newcommand{\Ptwoneone}{{$\text{P}_{20}\text{S}\cdot1.0\text{~IL}$}}
\newcommand{\Ptwonetwo}{{$\text{P}_{20}\text{S}\cdot2.0\text{~IL}$}}
\newcommand{\Ptwonefour}{{$\text{P}_{20}\text{S}\cdot4.0\text{~IL}$}}
\newcommand{\Ptonezero}{{$\text{P}_{10}\text{S}\cdot0.0\text{~IL}$}}
\newcommand{\Ptoneone}{{$\text{P}_{10}\text{S}\cdot1.0\text{~IL}$}}
\newcommand{\Ptonetwo}{{$\text{P}_{10}\text{S}\cdot2.0\text{~IL}$}}
\newcommand{\Pfoneone}{{$\text{P}_{5}\text{S}\cdot1.0\text{~IL}$}}

\newcommand{\LiOPEO}{{$\text{Li}^\text{+}-\text{O}_\text{PEO}$}}
\newcommand{\LiOTFSI}{{$\text{Li}^\text{+}-\text{O}_\text{TFSI}$}}
\newcommand{\LiNTFSI}{{$\text{Li}^\text{+}-\text{N}_\text{TFSI}$}}
\newcommand{\MPPYNTFSI}{{$\text{N}_\text{PYR$_\mathrm{13}$}-\text{N}_\text{TFSI}$}}
\newcommand{\MPPYOTFSI}{{$\text{N}_\text{PYR$_\mathrm{13}$}-\text{O}_\text{TFSI}$}}

\hyphenation{AMBER}

\begin{abstract}
We present an extensive molecular dynamics (MD) simulation study of the lithium ion transport in ternary polymer 
electrolytes consisting of poly(ethylene oxide) (PEO), lithium-bis\-(tri\-fluoro\-meth\-ane)sulfon\-imide (LiTFSI) and 
the ionic liquid {\it N}-methyl-{\it N}-propyl\-pyr\-rolid\-inium bis\-(tri\-fluoro\-methane)\-sulfon\-imide (PYR$_\mathrm{13}$TFSI). 
In particular, we focus on two different strategies by which the ternary electrolytes can be devised, namely by (a)~\emph{adding} 
the ionic liquid to \Ptwenty, and (b)~\emph{substituting} the PEO chains in \Ptwenty\ by the ionic liquid. 
In order to grasp the changes of the overall lithium transport mechanism, we employ an analytical, Rouse-based cation transport model 
(Maitra \etal, {\it Phys. Rev. Lett.}, {\bf 2007}, 98, 227802), which has originally been devised for binary PEO-based electrolytes. 
This model distinguishes three different microscopic transport mechanisms, each quantified by an individual time scale. 
In the course of our analysis, we extend this mathematical description to account for an entirely new transport mechanism, 
namely the TFSI-supported diffusion of lithium ions decoupled from the PEO chains, which emerges for certain stoichiometries. 
We find that the segmental mobility plays a decisive role in PEO-based polymer electrolytes. 
That is, whereas the addition of the ionic liquid to \Ptwenty\ plasticizes the polymer network and thus also increases the 
lithium diffusion, the amount of free, mobile ether oxygens reduces when substituting the PEO chains by the ionic liquid, 
which compensates the plasticizing effect. 
In total, our observations allow us to formulate some general principles about the lithium ion transport mechanism in ternary 
polymer electrolytes. 
Moreover, our insights also shed light on recent experimental observations (Joost \etal, {\it Electrochim. Acta}, {\bf 2012}, 86, 330). 
\end{abstract}



\maketitle

\section{Introduction}

\begin{figure*}
 \centering
 \subfigure[ \label{fig:snapshot_add} ]{ \includegraphics[scale=0.3]{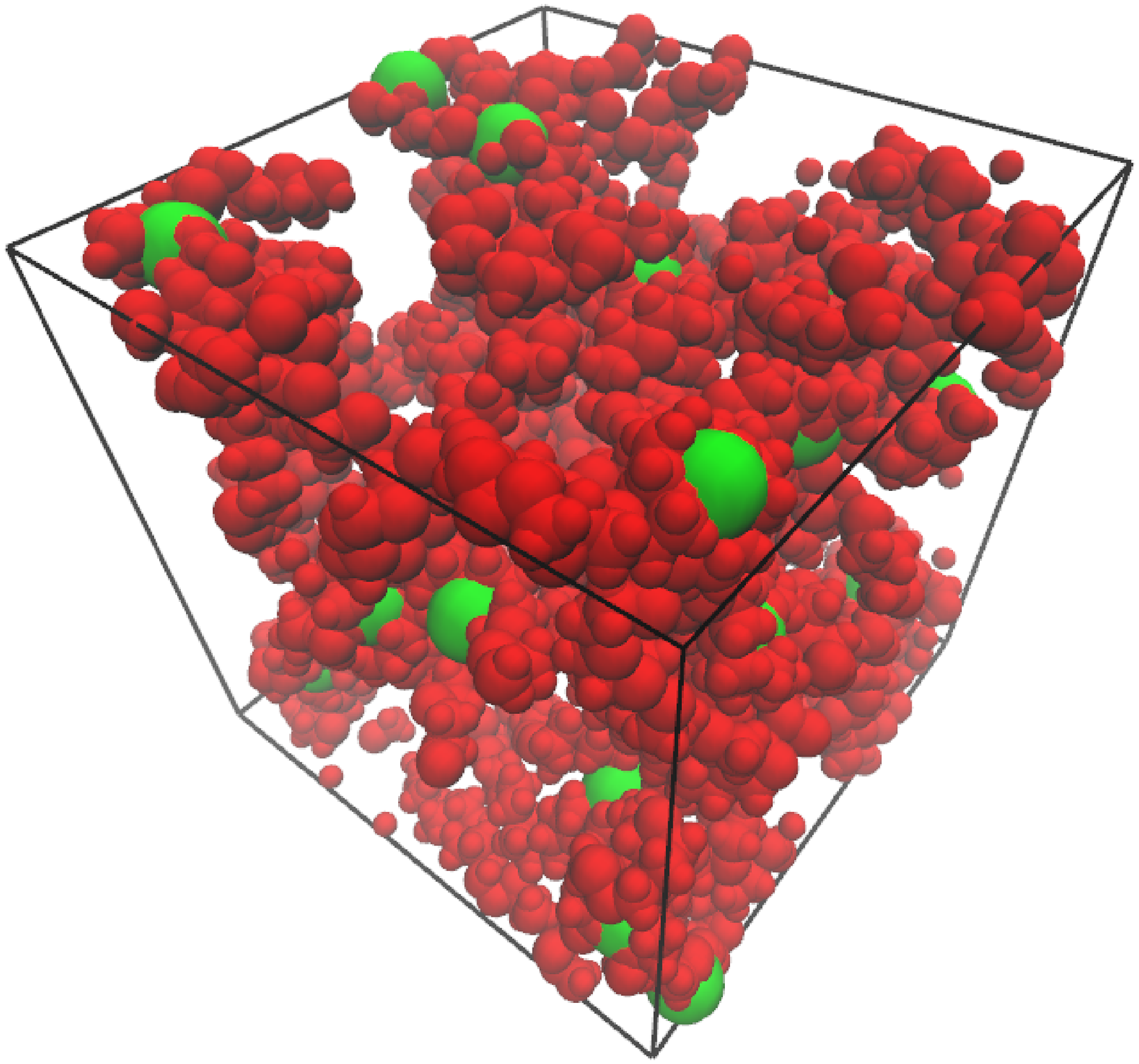} }
 \subfigure[ \label{fig:snapshot_sub} ]{ \includegraphics[scale=0.3]{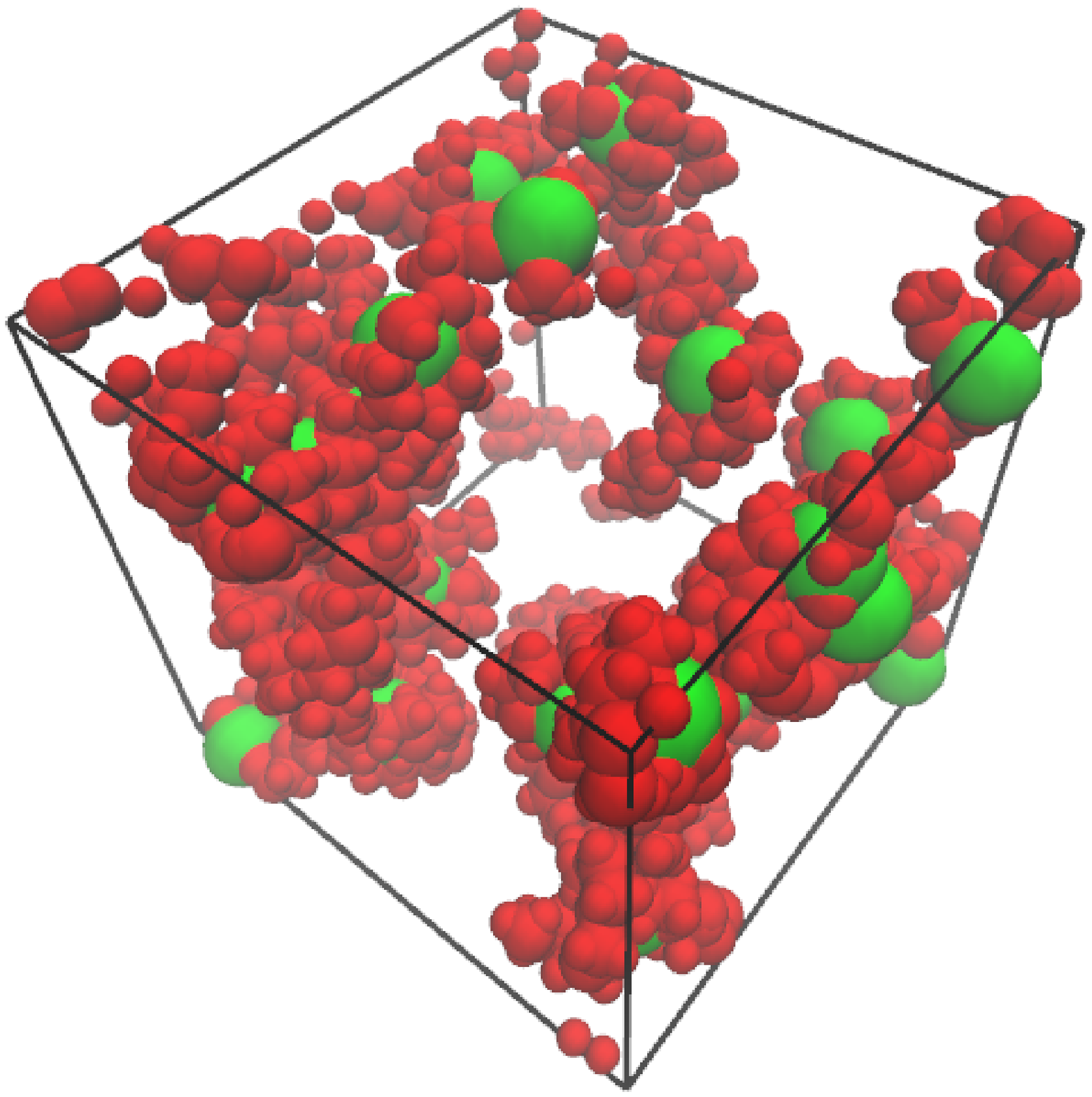} }
 \caption{Snapshots from the simulation showing (a)~the \emph{addition} of IL to the 
          binary polymer electrolyte \Ptwenty\ and (b)~the \emph{substitution} of PEO 
          chains by IL molecules from the same electrolyte. 
          PEO chains are shown in red, lithium ions in green, whereas all other 
          ions (PYR$_\mathrm{13}^+$ and TFSI$^-$) are invisible. }
 \label{fig:snapshots}
\end{figure*}

Polymer electrolytes -- typically consisting of an amorphous polymer matrix and a salt dissolved in it~\cite{Wright} -- 
are promising candidates for many technological devices such as batteries or fuel cells~\cite{Armand,Gray,BruceJCSFaraday1993}. 
Here, a commonly used polymer is poly(ethylene oxide) (PEO), whereas lithium-bis\-(tri\-fluoro\-meth\-ane)sulfon\-imide (LiTFSI) 
is often employed as conducting salt, as the large TFSI anion reduces ion aggregation. 
However, at ambient temperatures, the conductivity is still too low for an efficient technological use. 

Several attempts have been made in the past to overcome this deficiency. 
Common modifications of the classical polymer-salt systems are the 
addition of nanoparticles~\cite{ScrosatiSSI1992} or organic solvent 
molecules~\cite{BorghiniElectrochimActa1996,BandaraElectrochimActa1998,KimSSI2002}. 
However, in case of the latter, the resulting electrolytes 
suffer from the high volatility and thus flammability of 
the solvent as well as from its reaction with lithium metal 
electrodes~\cite{KimSSI2002}. 
Here, Passerini~\etal~\cite{PasseriniElectrochemCommun2003,PasseriniElectrochemActa2005,PasseriniJPowerSources2006,PasseriniJoost} 
demonstrated that the use of an ionic liquid (IL) instead of a conventional 
solvent has several advantages: 
The ILs are non-volatile, non-flammable~\cite{AdamNature2000} 
and exhibit a wide electrochemical stability window~\cite{MacFarlaneNature1999,MacFarlaneJPCB1999,PasseriniJPCB2008}. 
In this way, improved polymer electrolytes can be designed, 
which show an increased conductivity combined with inherent 
stability~\cite{PasseriniElectrochemCommun2003,PasseriniJoost}, 
and are thus ideal to create light-weighted but powerful 
batteries~\cite{PasseriniElectrochemActa2005,PasseriniJPowerSources2006}. 

In a recent molecular dynamics (MD) simulation study~\cite{DiddensMacroLett2013}, 
we investigated a ternary polymer electrolyte 
consisting of \Ptwenty\ and a variable mole fraction $x$ of the IL 
{\it N}-methyl-{\it N}-propyl\-pyr\-rolid\-inium bis(trifluoromethane)sulfonimide (PYR$_\mathrm{13}$TFSI). 
As observed experimentally by Passerini~\etal~\cite{PasseriniElectrochemCommun2003,PasseriniJoost}, we also 
found a clear increase of the lithium diffusion coefficient with $x$, and were able to show that the main 
reason for this enhancement can be attributed to the plasticization of the PEO backbone by the IL. 
That is, the presence of the IL enhances the PEO dynamics, and consequently, the lithium ions 
coordinated to the PEO backbone also become faster. 

In this contribution, we will significantly expand the scope of our previous analysis~\cite{DiddensMacroLett2013}, 
and investigate the lithium transport in ternary polymer electrolytes in a more general way. 
This is not only due to the fact that we present additional insights on the ternary \PxIL\ mixtures (in which IL is \emph{added} 
to binary \Ptwenty), but also follow a conceptually different approach how to create ternary polymer electrolytes, that is, we also 
\emph{substitute} PEO chains in \Ptwenty\ by PYR$_\mathrm{13}$TFSI molecules. 
This procedure is carried out in such a way that the overall lithium volume concentration is held constant. 
Figure~\ref{fig:snapshots} shows snapshots from the simulations for the two different scenarios. 
The motivation for our extended study is twofold: 
First, although Passerini~\etal\ found in their recent work~\cite{PasseriniJoost} that the lithium diffusion coefficient 
increases with the IL fraction for an ether oxygen-to-lithium ion ($\text{EO}:\text{Li}$) ratio of $20:1$, they simultaneously 
observed no significant changes for lower ratios of $10:1$ or $5:1$. 
This observation becomes even more puzzling by additional insights from Raman measurements~\cite{PasseriniJoost}, which revealed 
that with increasing $x$, the lithium-TFSI coordination increases at the expense of the lithium-PEO coordination (especially for 
low $\text{EO}:\text{Li}$ ratios). 
Although these findings rather suggest a lithium transport mechanism which is at least partly decoupled from the PEO chains, it is 
obvious from the approximately constant lithium diffusion coefficients that the lithium motion cannot be similar to that in pure IL/LiTFSI 
electrolytes, which would be significantly faster~\cite{PasseriniJPCB2005,PasseriniJPCC2011}. 
Nevertheless, the transport mechanism has to change at some point when PEO is successively replaced by PYR$_\mathrm{13}$TFSI 
molecules, which gives us a second, theoretical motivation for our extended study. 
Thus, the concentration series of the PEO substitution serves as a model electrolyte in a double sense, as both the 
$\text{EO}:\text{Li}$ ratio decreases with increasing IL fraction, while at the same time a crossover from a PEO-based lithium 
transport mechanism to a TFSI-based transport mechanism is enforced. 

\begin{figure}
 \centering
 \includegraphics[scale=0.45]{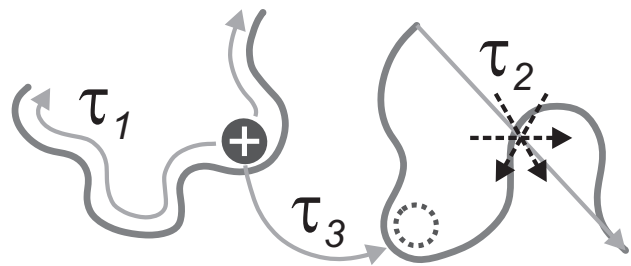}
 \caption{Sketch depicting the three different cation 
          transport mechanisms in PEO-salt electrolytes. 
          Each mechanism is characterized by a specific 
          time scale. }
 \label{fig:sketch_transport_processes}
\end{figure}

In addition to the MD simulations, we also employ an analytical ion transport model~\cite{MaitraHeuerPRL2007} to clarify the issues 
outlined above. 
This model expresses the macroscopic lithium diffusion in PEO-based polymer electrolytes via three different microscopic mechanisms, 
each characterized by a specific time scale (see sketch in Figure~\ref{fig:sketch_transport_processes}): 
1.~Diffusion along the PEO chain, which can be interpreted as an effective one-dimensional random walk along 
the curvilinear path of the polymer chain. 
This motion can be characterized by the time scale $\tau_1$ the ion needs to explore the entire PEO chain. 
2.~Segmental motion of the PEO chain, which can be separated into the center-of-mass motion and the internal 
dynamics. 
For non-entangled chains, the internal dynamics can be described by the Rouse model~\cite{RouseJCP1953} via an effective Rouse time 
$\tau_2$, characterizing the motion of the bound PEO segments. 
3.~Intersegmental transfer of the cation from one PEO chain to another, which can be quantified by the average 
residence time $\tau_3$ at a given chain. 
Previous simulations~\cite{MaitraHeuerPRL2007} have shown that this mechanism can be viewed as a renewal 
process within the framework of the Dynamic Bond Percolation (DBP) model~\cite{rev_DBP}, since the dynamics of 
a given lithium ion becomes independent of its past after being transferred to another PEO chain. 
One particular advantage of our model is that it allows one to extrapolate the lithium diffusion coefficients 
derived from the numerical data to the experimentally relevant long-chain limit. 

However, one should keep in mind that the ion transport model has originally been developed for the archetypal, binary 
polymer electrolytes. 
Although our previous study~\cite{DiddensMacroLett2013} showed that the lithium transport mechanism qualitatively 
remains the same for the case of IL addition, it is a priori unclear if this also holds for the PEO substitution, 
as the transport mechanism will certainly change at some point when PEO is removed from the electrolyte. 
In fact, it will turn out that the IL-mediated lithium transport may substantially contribute to the overall 
lithium migration, and we will propose a revised model accounting for this additional, fourth transport mechanism. 

Further advancements compared to our original study~\cite{DiddensMacroLett2013} are comprehensive analyses related 
to the microscopic factors contributing to the individual transport mechanisms. 
For example, we will demonstrate that the plasticization by the IL is only one important requirement to yield an 
enhanced segmental mobility and thus a faster lithium diffusion. 
Moreover, the effect of the IL on the polymer dynamics is also interesting from a fundamental point of view, as it 
gives rise to hydrodynamic interactions among distinct PEO segments, reflected by a Zimm-like scaling of the Rouse-mode 
relaxation times. 
Concerning the other mechanisms, we will show that \eg the renewal process (characterized by $\tau_3$) can be understood 
in terms of a few structural and dynamical observables. 
These additional insights could ultimately contribute to a more directed improvement of battery materials. 

For all ternary electrolytes, \Ptwenty\ will be used as a reference. 
In what follows, the electrolytes with a constant $\text{EO}:\text{Li}$ ratio, \ie $20:1$, and a variable amount $x$ of PYR$_\mathrm{13}$TFSI 
will be denoted as \PxIL. 
The new electrolyte types, devised to study the differences in the lithium transport mechanisms between 
binary PEO/LiTFSI and PYR$_\mathrm{13}$TFSI/LiTFSI, will be denoted as \PminusxIL. 
Here, $\alpha$ is the ratio of the respective partial molar volumes of a PYR$_\mathrm{13}$TFSI ion pair and 
a PEO monomer (see below). 

For reasons of simplicity, we also abbreviate PEO as \lq P\rq\ (\ie polymer or PEO) and LiTFSI 
as \lq S\rq\ (\ie salt) in the following, leading to the short-hand notations \Px\ and \Pminusx\ 
for the two distinct classes of ternary electrolytes. 

This article is organized as follows: 
Section~\ref{sec:md_details} describes the details of the MD simulation procedure and the setup of the systems. 
In Section~\ref{sec:structure}, we discuss the lithium ion coordination sphere and the structure of the PEO chains. 
Section~\ref{sec:lithium_dynamics} deals with the two transport processes involving changes in the lithium coordination 
sphere (intersegmental transfer and diffusion along the chain), followed by a detailed analysis of the polymer dynamics 
(Section~\ref{sec:polymer_dynamics}). 
Having characterized all three transport mechanisms, we compare the predictions of our model with the experimental data from 
Passerini~\etal~\cite{PasseriniJoost} (Section~\ref{sec:cmp_exp}). 
Finally, we conclude (Section~\ref{sec:conclusion}).

\section{MD Simulation Methodology}
\label{sec:md_details}

The MD simulations were performed with a modified version of the sander module of the AMBER 
package~\cite{AMBER10}, allowing us to use a force field specifically designed for 
PEO/LiTFSI~\cite{BorodinJPCB2006_PEO,BorodinJPCB2006_PEOLiTFSI} and PYR$_\mathrm{13}$TFSI~\cite{BorodinJPCB2006_IL} 
(see appendix, Section~\ref{sec:validation} for a validation of the modified AMBER code and Section~\ref{sec:benchmark} 
for a benchmark assessing its performance).  

The initial configurations were created by randomly placing the individual molecules on a simple 
cubic lattice with a lattice constant of $36\text{~\AA}$ irrespective of their type, mixing the 
system in this way. 
In case of PEO, the chains already had coiled conformations as under melt conditions, thereby 
circumventing an expensive equilibration. 

The binary PEO/LiTFSI electrolyte contained $10$~PEO chains with $N=54$ monomers each as well as 
$27$~LiTFSI molecules. 
In case of the addition of IL, \ie \Px, the simulation cell also contained $18$ ($x=0.66$) or $87$ 
($x=3.24$) PYR$_\mathrm{13}$TFSI molecules. 
The latter system is shown in Figure~\ref{fig:snapshot_add}. 
In case of the substitution of PEO (\Pminusx) and thus a constant lithium volume concentration, the 
partial molar volumes of PEO and PYR$_\mathrm{13}$TFSI must be known. 
Through several short equilibration runs of the respective binary electrolytes, namely $\text{PEO}_n\text{LiTFSI}$ 
($n=30$, $20$ and $10$) and $\text{PYR}_\mathrm{13}\text{TFSI}\cdot x\text{~LiTFSI}$ 
($x=0.20$, $0.24$ and $0.26$) in the $NpT$ ensemble, we found that the molar volumes of both PEO and PYR$_\mathrm{13}$TFSI 
are independent of the LiTFSI fraction for the investigated concentration ranges. 
Their ratio reveals that approximately $\alpha=7.14$ PEO monomers occupy the same volume as one PYR$_\mathrm{13}$TFSI 
ion pair. 
Based on this value, we created the following systems: $8$~PEO chains with $15$~PYR$_\mathrm{13}$TFSI molecules 
(\Psixt), $6$~PEO chains with $30$~PYR$_\mathrm{13}$TFSI molecules (\Ptwel), and $4$~PEO chains with 
$45$~PYR$_\mathrm{13}$TFSI molecules (\Pei, see Figure~\ref{fig:snapshot_sub}), all of them containing, as above, $27$~LiTFSI ion pairs. 
Neat PEO ($16$ chains with $N=54$) and the pure PYR$_\mathrm{13}$TFSI/LiTFSI electrolyte ($76$~PYR$_\mathrm{13}$TFSI 
molecules and $27$~LiTFSI ion pairs, \Pzero) serve as reference within this context. 

The systems were equilibrated in the $NpT$ ensemble for $70-80\text{~ns}$ using the PME technique~\cite{DardenJCP2000}. 
Afterwards, production runs with a total length of $200\text{~ns}$ were performed in the $NVT$ ensemble. 
A time step of $1\text{~fs}$ was used in all simulations to propagate the systems. 
The temperature was maintained by the Berendsen thermostat~\cite{BerendsenJCP1984} with a reference temperature of 
$423\text{~K}$. 
All bonds involving hydrogen were constrained by the SHAKE algorithm~\cite{SHAKE}. 
The induceable point dipoles were integrated by a Car-Parrinello-like 
scheme~\cite{vanBelleMolPhys1992}, and the charge-dipole interactions between atoms 
separated by three bonds ($1$-$4$ interactions) were scaled by a factor of $0.2$. 
For all other nonbonded contributions, the full $1$-$4$ interaction was taken 
into account. 
Dipole-dipole interactions were damped using a Thole screening~\cite{TholeChemPhys1981} with a dimensionless damping 
parameter~\cite{BorodinJPCB2006_PEO} of $a_\mathrm{T}=0.4$. 

\begin{table*}
 \centering
 \footnotesize
 \begin{tabular}{l c c r c c}
  \hline
  \hline
  system & $r_1$ & $r_2$ & $b_0^2$ [$\text{\AA$^2$}$] & $\langle{\bf R}_\mathrm{e}^2\rangle$ [$\text{\AA$^2$}$] & $\langle{\bf R}_\mathrm{g}^2\rangle$ [$\text{\AA$^2$}$] \\
  \hline
  PEO & - & - & $10.4$ & $1904\pm43$ & $306\pm35$ \\
  \hline
  \Ptwen\ & $0.47$ & $0.53$ & $9.7$ & $1573\pm112$ & $260\pm42$ \\
  \hline
  \multicolumn{6}{c}{ \Px\ } \\
  \hline
  \Psome\ & $0.53$ & $0.47$ & $9.7$ & $1654\pm121$ & $272\pm38$ \\
  \Pmore\ & $0.76$ & $0.24$ & $9.6$ & $1498\pm57$ & $249\pm37$ \\
  \hline
  \multicolumn{6}{c}{ \Pminusx\ } \\
  \hline
  \Psixt\ & $0.55$ & $0.45$ & $9.5$ & $1767\pm271$ & $267\pm50$ \\
  \Ptwel\ & $0.64$ & $0.35$ & $9.3$ & $1349\pm150$ & $227\pm43$ \\
  \Pei\ & $0.84$ & $0.14$ & $9.1$ & $1840\pm430$ & $280\pm80$ \\
  \hline
  \hline
 \end{tabular}
 \normalsize
 \caption{Fraction of lithium ions coordinating to one ($r_1$) or two ($r_2$) PEO chains, 
          mean squared chemical bond length $b_0^2$, mean squared end-to-end 
          vector $\langle{\bf R}_\mathrm{e}^2\rangle$ and mean squared radius 
          of gyration $\langle{\bf R}_\mathrm{g}^2\rangle$ of the PEO chains. }
 \label{tab:structure}
\end{table*}

Table~\ref{tab:boxlength} (appendix) shows the average values of the box 
lengths as determined from the $NpT$ runs and subsequently used in the $NVT$ simulations. 
One indeed observes that box length is nearly the same for all \Pminusx\ systems. 
Slight deviations for \Pzero\ may result from rounded molecule numbers in the simulation.

\section{Structural Properties}
\label{sec:structure}

\subsection{Lithium Coordination}

In order to quantify the local structure around the lithium ions, radial distribution functions 
(denoted as $g(r)$ in the following) have been computed for the atom pairs \Li$-\text{O}_\text{PEO}$ 
and \Li$-\text{O}_\text{TFSI}$ (see \ref{ssec:gofr}). 
Both coordination types exhibit a sharp peak around $2\text{~\AA}$, corresponding to the first 
coordination shell, which is in good agreement with neutron diffraction experiments~\cite{MaoPRL2000} 
and quantum chemistry calculations~\cite{JohanssonPolym1999,BaboulJAmChemSoc1999}. 
At larger distances no significant structural arrangement can be found. 
When successively adding the IL, one observes that the peak positions of both \LiOPEO\ 
and \LiOTFSI\ remain the same for all electrolytes. 
Thus, the same criterion to define temporary lithium bonds will be used for the subsequent analysis. 
We consider a EO and a \Li\ as bound if their distance is not larger than $3.0\text{~\AA}$. 
In analogy, we consider a \Li\ and a TFSI oxygen as bound if their distance is not larger than $2.7\text{~\AA}$. 

The probability distribution functions to find a lithium ion with a certain number of EOs 
or TFSI oxygens in its first coordination shell is also shown in the appendix (Figure~\ref{fig:coordbins}). 
To summarize, we find that the predominant lithium coordination consists of $4-5$ EOs for 
all electrolytes, which is again in good agreement with the experimental data~\cite{MaoPRL2000} and 
the quantum-chemical results~\cite{JohanssonPolym1999,BaboulJAmChemSoc1999}. 
In those complexes where the $4-5$ EOs originate from a single PEO molecule, the polymer 
chain wraps helically around the cation. 
For complexes involving two PEO chains, typically $2-3$ EOs from each chain coordinate to 
the ion. 

Coordinations by TFSI oxygens are mostly rare, and often the anion coordinates only briefly 
to the lithium ion (cf. Figure~\ref{fig:snapshots}, where all ions are in the vicinity of a PEO chain). 
The only exceptions are \Pei\ and, though less pronounced, \Ptwel. 
Here, the lithium ions more likely coordinate to $1-2$ TFSI oxygens in contrast to all other systems, 
and some are even coordinated to TFSI only ($0.4\text{~\%}$ for \Ptwel\ and $2.5\text{~\%}$ for \Pei). 
As in pure IL/LiTFSI, the prevalent coordination number in the latter scenario is about $3-4$ TFSI oxygens, 
originating mostly from different anions. 
This specific coordination has also been observed previously in MD simulations~\cite{BorodinJPCB2006_ILLiTFSI}, 
although experimental work emphasizes that also the $\text{Li}(\text{TFSI})_2$ complex is important~\cite{LasseguesPCCP2006}. 

Table~\ref{tab:structure} summarizes the fraction of lithium ions coordinating to one or two PEO chains 
(denoted as $r_1$ and $r_2$, respectively). 
Coordinations to three PEO chains were rarely observed and had a very brief life time of a few picoseconds only. 
Therefore, these events were neglected for the subsequent analysis. 
For the pure polymer electrolyte, the fractions of complexes involving one and two chains are nearly equal. 
Similar binding energies for both coordination types have also been found in quantum chemistry 
calculations~\cite{BaboulJAmChemSoc1999}. 
The fraction of lithium ions coordinating to one PEO molecule increases with the IL content for both types of electrolytes. 

In case of \Px, this is a consequence of the reduced PEO concentration, as it becomes less likely that a lithium ion encounters a 
second chain in the semidilute case. This is also confirmed by the observation that the fraction of \Li\ coordinating to one PEO 
chain increases linearly with $x$, showing that the changes in the coordination sphere are purely statistical. 

For \Pminusx, however, these trends are stronger than linear, which can be attributed to the fact that the PEO chains are 
successively removed from the system, and the remaining PEO chains become more crowded by \Li\ (see below). 
Thus, the different coordination shell can not solely be explained as a simple dilution effect.

\subsection{Statical Polymer Structure}

Due to the helical coordination structure of the PEO backbone, the local polymer 
structure changes, and the conformational phase space of the chain is reduced. 
Moreover, in case of the ternary electrolytes, the additional IL molecules  
dilute the PEO chains, thus inducing a crossover from a polymer melt 
to a semidilute solution, which may also alter the equilibrium conformation 
of the polymer chains~\cite{DoiEdwards}. 
Table~\ref{tab:structure} summarizes the mean squared distance $b_0^2$ 
between two chemical monomers and the mean squared end-to-end vector 
$\langle{\bf R}_\mathrm{e}^2\rangle$. 
Due to the crown-ether-like coordination of the PEO backbone, $b_0^2$ is 
smaller in all lithium-containing systems. 
For \Pminusx, this trend becomes more pronounced with increasing IL concentration. 

A contraction of the polymer chain can also be observed for \Px\ from 
$\langle{\bf R}_\mathrm{e}^2\rangle$, which also decreases, whereas for 
\Pminusx\ no clear predictions can be made within the statistical uncertainties. 
When determining the ratio of $\langle{\bf R}_\mathrm{e}^2\rangle$ and 
the radius of gyration $\langle{\bf R}_\mathrm{g}^2\rangle$ (Table~\ref{tab:structure}), one finds 
values close to the ideal ratio of six for a Gaussian chain~\cite{DoiEdwards} 
for all electrolytes, again indicating that on a global scale the chains are approximately Gaussian. 

\begin{figure}
 \centering
 \includegraphics[scale=0.3]{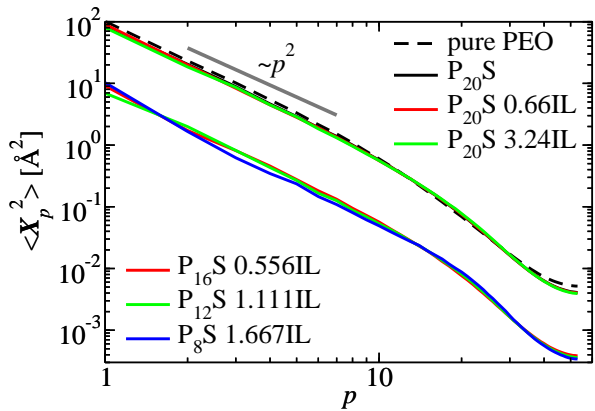}
 \caption{Rouse mode amplitudes $\langle{\bf X}_p^2\rangle$ in dependence of 
          the mode number $p$ calculated from the expression of the discrete 
          Rouse model, \ie ${\bf X}_p(t) = (1/N)\sum_{n=1}^N \cos((p\pi (n-1/2)/N)){\bf R}_n(t)$. 
          The curves of the \Pminusx\ electrolytes have been shifted by one order of magnitude. }
 \label{fig:Xp2}
\end{figure}

Also for the scaling of the Rouse mode amplitudes $\langle{\bf X}_p^2\rangle$, 
we observe only slight deviations from the respective curve for pure PEO (Figure~\ref{fig:Xp2}). 
Here, the modes ${\bf X}_p$ were determined from the simulation data using the respective 
expression of the discrete Rouse model~\cite{VerdierJCP1966}, 
\ie ${\bf X}_p(t) = (1/N)\sum_{n=1}^N \cos(p\pi(n-1/2)/N)\,{\bf R}_n(t)$, where the ${\bf R}_n$ 
denote the position vectors of the individual monomers.  
In the limit of low mode numbers $p$, we find the expected Rousean scaling 
$\langle{\bf X}_p^2\rangle\propto p^{-2}$, again demonstrating that the 
chain structure remains relatively ideal upon the addition of IL. 
Thus, no significant swelling of the chains can be observed, and the 
structural properties are similar as in neat PEO. 
On a local scale however, the PEO chains become rather contracted due to 
the helical coordination sphere of the lithium ions.

\section{Lithium Dynamics}
\label{sec:lithium_dynamics}

\begin{figure}
 \centering
 \includegraphics[scale=0.3]{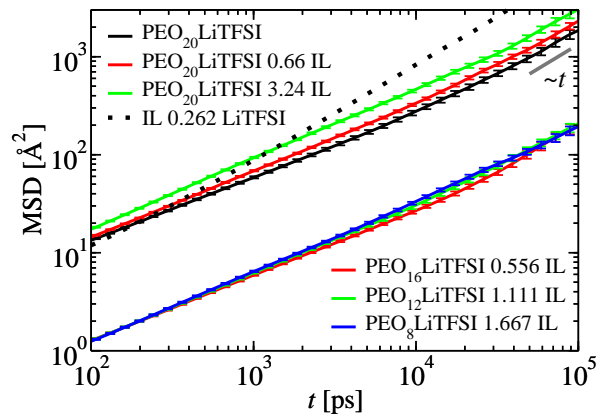}
 \caption{MSDs of the lithium ions in the individual electrolytes. 
          The curves of the \Pminusx\ electrolytes have been shifted by one order of magnitude. 
          The curve for the pure IL-salt solution \Pzero\ is also shown for comparison. }
 \label{fig:msd_Li}
\end{figure}

The MSDs of the lithium ions in the individual electrolytes are shown in Figure~\ref{fig:msd_Li}. 
For \Px, one indeed observes an increase of the lithium mobility upon the addition of IL. 
In particular, this increase becomes visible at $1-10\text{~ns}$, whereas the onset to 
diffusive behavior occurs on the same time scale for all systems. 
Nevertheless, the MSDs are much smaller than for pure \Pzero, and the crossover to diffusion 
occurs much later. 
This is consistent with the observation that nearly all cations coordinate to the PEO chains, 
and that the lithium transport in the ternary electrolytes changes not too drastically as compared 
to \Ptwen. 
Interestingly, for \Pminusx, no significant difference in the lithium MSD can be found. 

In the following, we will employ the lithium ion transport model in order to rationalize these observations. 
First, we will focus on the intersegmental transfer and the diffusion along the chain, followed by an in-detail 
analysis of the polymer dynamics. 
Once all ingredients are combined, we will use the transport model to compute the macroscopic lithium diffusion 
coefficients from the numerical data, and compare these values to the experimental data from Passerini~\etal~\cite{PasseriniJoost}.

\subsection{Lithium Ion Transfer Mechanism}

First, we will study the lithium ion transfer between two different polymer chains, which is vital for the long-range cation transport and thus any 
macroscopically measurable lithium diffusion. 
That is, when only considering the diffusion along the chain and the PEO dynamics as possible transport mechanisms, the ion remains 
confined to a finite volume characterized by the radius of gyration of the PEO chains. 
Moreover, the diffusion of the center of mass of the polymer chain becomes irrelevant in the experimentally relevant long-chain limit. 
Previous MD simulations focusing on this mechanism have shown that the dynamics of the lithium ion is independent from its past 
after such a transfer processes~\cite{MaitraHeuerPRL2007}, leading to Markovian behavior for time scales larger than $\tau_3$, and that 
these processes can thus be regarded as renewal events in the spirit of the DBP model~\cite{rev_DBP}. 

We start with the investigation of the detailed mechanism of the cation transfer process. 
During most transfers, the ion is coordinated to two PEO chains simultaneously (\ie the 
leaving chain and the entering chain) as a transition state. 
In some events, the cation is also temporarily coordinated by TFSI anions only and migrates 
to another PEO chain in this way, which will be discussed below. 

\begin{figure}
 \centering
 \includegraphics[scale=0.3]{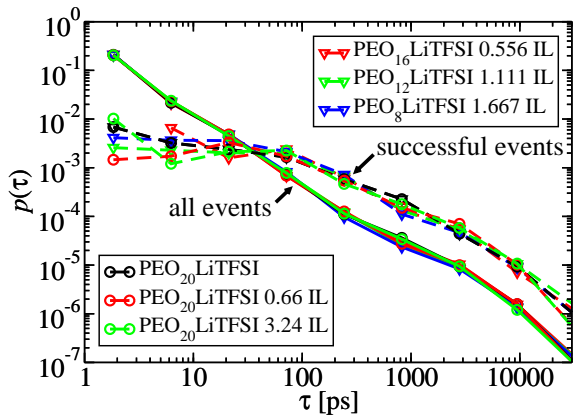}
 \caption{Histogram of the life times $\tau$ of the intermediate complexes involving two PEO chains 
          which facilitate the cationic transfer. Solid: all events, dashed: successful events only 
          (see text). }
 \label{fig:tau_trans}
\end{figure}

Figure~\ref{fig:tau_trans} shows a histogram of the life times $\tau$ of the intermediate in which a 
\Li\ is coordinated by two PEO chains (solid lines). 
One clearly observes that the resulting distribution is very broad. 
Whereas most complexes exist only briefly (\ie a few picoseconds), some exist over nearly the 
whole simulation length of $200\text{~ns}$. 
For all electrolytes, no significant difference can be found for $p(\tau)$, suggesting a rather 
universal mechanism. 
This is also reasonable when keeping in mind that the lithium coordination sphere containing two 
PEO chains is relatively compact, and that the surrounding molecules thus hardly affect the 
transfer process once the complex has formed. 

Of course, it is questionable if also the brief coordinations in Figure~\ref{fig:tau_trans} can result 
in a successful cation transfer, since the separation of the leaving chain and the simultaneous 
formation of a new, stable coordination sphere on the entering chain will need a certain time. 
For this purpose, Figure~\ref{fig:tau_trans} also shows $p(\tau)$ for complexes which resulted in a 
successful transfer. 
The criterion for a successful transfer was that the lithium ion remained detached from the old 
chain for at least $100\text{~ps}$, which was motivated by a more detailed analysis revealing that non-Markovian, 
short-time backjumps to the old chain occur up to about $100\text{~ps}$ (not shown). 
From Figure~\ref{fig:tau_trans} one indeed observes that the probability for short $\tau$ is lower 
for real renewal events (dashed lines). 
Again, the curves are nearly identical for all electrolytes. 

In order to study the transfer mechanism in more detail, we monitored the progress of the cation 
transfer as follows: When a second PEO chain enters the coordination sphere of a cation that is 
already coordinated to a PEO molecule, the EO coordination number $n$ at the first chain naturally 
decreases. 
For brief coordinations of the second chain, the crown-ether-like structure of the first chain 
will hardly be perturbed. 
On the other hand, for longer times, one would expect a rather symmetric coordination by both PEO 
chains, leading to smaller $n$. 
In case of a successful cation transfer, the EO coordination number decreases even further to zero. 
The minimum coordination number $n_\mathrm{min}$ that is reached during $\tau$ can therefore be 
used to monitor the progress and the success of the cation transfer. 

\begin{figure}
 \centering
 \includegraphics[scale=0.3]{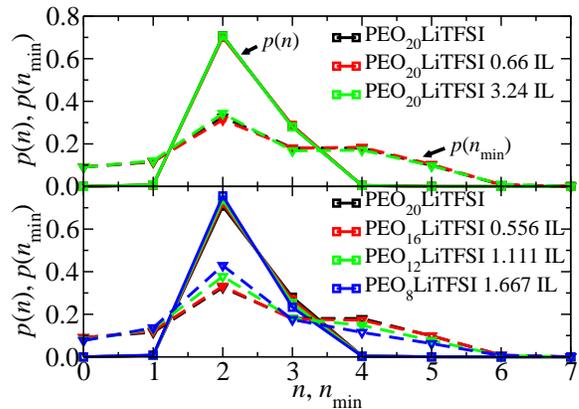}
 \caption{Probability $p(n)$ to find a specific EO coordination number $n$ for a cation coordinated 
          to two PEO chains (solid) as well as the probability $p(n_\mathrm{min})$ to find a minimum 
          EO coordination number $n_\mathrm{min}$ on the first PEO chain during the life time of the 
          complex (dashed). }
 \label{fig:p_n_min}
\end{figure}

Figure~\ref{fig:p_n_min} shows the probability $p(n_\mathrm{min})$ to find a specific minimum coordination 
number $n_\mathrm{min}$ as well as the probability to find any coordination number $n$ on the first 
PEO chain during $\tau$. 
One clearly observes that on average $2-3$~EOs of each chain coordinate to the \Li (solid lines). 
This coordination is independent from the IL concentration as expected from the dense packing of the 
coordination sphere. 
For $p(n_\mathrm{min})$, one notices that the brief coordinations corresponding to $n_\mathrm{min}=4-5$ 
become less likely with increasing IL concentration for \Pminusx. 
As a consequence, the relative probability that a complex with $n_\mathrm{min}=2$ is formed increases. 
For \Px, no such effect can be observed, and $p(n_\mathrm{min})$ remains basically unaltered by the 
presence of the IL. 
This indicates that the increased probability of $n_\mathrm{min}=2$ for \Pminusx\ is not related to a 
simple concentration effect, which would also be present for \Px, but rather to the fact that the PEO molecules 
are successively substituted by IL. 
Due to this substitution, each PEO chain coordinates to more cations, resulting in an enhanced rigidity of the PEO backbone, 
which impedes the formation of the helical structure with $n=4-5$ and thus leads to a larger value for $p(n_\mathrm{min}=2)$. 

In total, a complete cation transfer with $n_\mathrm{min}=0$ has the same relative probability of about 
$10\text{~\%}$ for all IL concentrations and $\text{EO}:\text{Li}$ ratios. 
Obviously, the critical step is the encounter of a second PEO chain, or, more precisely, a free PEO segment. 
Once the complex has formed, it decays according to a universal distribution of life times (Figure~\ref{fig:tau_trans}). 
A remarkable feature, however, is the fact that although $p(n_\mathrm{min}=2)$ (which corresponds to a symmetric coordination 
by both PEO chains) for \Pminusx\ increases with $x$, the complete transfer with $p(n_\mathrm{min}=0)$ does not become more 
likely, indicating that other, less trivial factors affect the outcome of a PEO-\Li-PEO encounter. 

It is also worth noting that, in contrast to earlier findings from MD simulations of a PEO/LiBF$_4$ 
electrolyte~\cite{MaitraHeuerPRL2007}, no significant influence of the anion on the  PEO-to-PEO transfer 
could be observed. 
Most likely, the minor importance of the TFSI anions for the transfer arises from the suppressed tendency 
to form ion pairs or higher-order ionic clusters as compared to BF$_4^-$.

\subsection{Renewal Time}

In order to determine $\tau_3$, we counted the number of transfer processes 
$N_\mathrm{trans}$ from the simulations. 
As above, brief transfers and successive backjumps after less than $100\text{~ps}$ 
were excluded, while on the other hand transfer processes via the anions were also 
counted, since they serve as a renewal process in the same way. 
Subsequently, the $\tau_3$-values were determined according to 
$\tau_3=t_\mathrm{max}N_{\mathrm{Li}^+}/N_\mathrm{trans}$, where $t_\mathrm{max}=200\text{~ns}$ 
is the simulation length and $N_{\mathrm{Li}^+}=27$ is the number of lithium ions in the 
simulation box. 

\begin{table*}
 \centering
 \footnotesize
 \begin{tabular}{l c c r r r r r}
  \hline
  \hline
  system & $\tau_3$ [ns] & $V_\mathrm{PEO}/V$ & $p_\mathrm{EO}^\mathrm{free}$ [\%] & $p_\mathrm{IL}$ [\%] & $p_\mathrm{trans,IL}$ [\%] & $\langle\tau_\mathrm{Li}\rangle_\mathrm{IL}$ [ns] & $\langle\Delta{\bf R}^2_\mathrm{Li}\rangle_\mathrm{IL}$ [$\text{\AA$^2$}$] \\
  \hline
  \Ptwen\ & $17.1\pm1.3$ & $0.87$ & $72.3$ & $0.02$ & $2.5$ & $0.25 \pm 0.08$ & $31 \pm 8$ \\
  \hline
  \multicolumn{8}{c}{ \Px\ } \\
  \hline
  \Psome\ & $18.4\pm1.4$ & $0.72$ & $71.9$ & $0.03$ & $1.0$ & $0.58 \pm 0.31$ & $34 \pm 13$ \\
  \Pmore\ & $24.1\pm1.3$ & $0.43$ & $73.1$ & $0.36$ & $8.5$ & $0.99 \pm 0.22$ & $102 \pm 26$ \\
  \hline
  \multicolumn{8}{c}{ \Pminusx\ } \\
  \hline
  \Psixt\ & $25.0\pm2.0$ & $0.70$ & $65.2$ & $0.07$ & $4.6$ & $0.38 \pm 0.14$ & $34 \pm 10$ \\
  \Ptwel\ & $28.4\pm2.2$ & $0.52$ & $54.7$ & $0.43$ & $13.7$ & $0.88 \pm 0.20$ & $79 \pm 17$ \\
  \Pei\ & $32.7\pm2.5$ & $0.35$ & $36.7$ & $2.50$ & $38.2$ & $2.10 \pm 0.31$ & $226 \pm 51$ \\
  \hline
  \hline
 \end{tabular}
 \normalsize
 \caption{Average residence times $\tau_3$ of a lithium ion at a given PEO chain. 
          Here, $p_\mathrm{trans,IL}$ is the probability that the lithium ion is transferred 
          by TFSI anions, $\langle\tau_\mathrm{Li}\rangle_\mathrm{IL}$ corresponds to the average 
          duration the \Li\ is coordinated to TFSI only in such a transfer, and 
          $\langle\Delta{\bf R}^2_\mathrm{Li}\rangle_\mathrm{IL}$ is the average squared 
          distance the ion covers during this time. }
 \label{tab:tau3}
\end{table*}

The individual values for $\tau_3$ are summarized in Table~\ref{tab:tau3}. 
One observes for both the \Px\ and the \Pminusx\ electrolytes that $\tau_3$ increases with increasing IL concentration. 
Since the PEO molecules become more and more diluted, this can at least partly be explained as a concentration effect. 
However, the stronger increase for \Pminusx\ indicates that not only the reduced PEO concentration, but also the reduced 
number of free EOs, which are necessary for a successful transfer, leads to an increase of $\tau_3$. 
To understand these trends in more detail, we will figure out the individual contributions to the $\tau_3$-values in the following. 
When the lithium ion moves through the electrolyte, it has to encounter a free PEO segment in order to be transferred (naturally, 
above the glass-transition temperature, the segments themselves also show significant motion). 
The probability for such an event can be estimated on the basis a few structural parameters: 
First, the volume element into which the \Li\ moves has to be occupied by PEO. 
Here, the probability to encounter any PEO segment (free or bound) can be obtained from the volume fraction of PEO in 
the simulation cell, \ie $V_\mathrm{PEO}/V$. 
Second, the PEO segment has to be free, which is necessary to form the helical coordination sphere. 
Here, we defined a free segment as a block of at least four free EOs, which occurs with probability $p_\mathrm{EO}^\mathrm{free}$. 
The overall probability to encounter a free PEO segment (at least four EOs) is then given by the product of $V_\mathrm{PEO}/V$ 
and $p_\mathrm{EO}^\mathrm{free}$ (both quantities are shown in Table~\ref{tab:tau3}). 
Apart from these structural ingredients, one also needs dynamical information how fast the PEO segments (and the attached lithium ions) move. 
That is, the faster both the lithium ions and the PEO segments rearrange, the more opportunities the ions have to find a free PEO segment 
and to be transferred. 
Although the effect of the IL on the polymer dynamics will be discussed in Section~\ref{sec:polymer_dynamics}, we nevertheless use the parameter 
$C_\mathrm{R}$ (Table~\ref{tab:tau12}), which is the prefactor of the Rouse-like $\text{MSD}\propto t^{1/2}$ for the average EOs, as an anticipatory 
measure for the polymer dynamics. 
Since $C_\mathrm{R}$ quantifies the motion of the average EOs, it implicitly contains the dynamics of both the free PEO segments and the lithium 
ions (due to the cooperative motion with the bound EOs, which are also contained in $C_\mathrm{R}$; see below). 
When plotting the product $(V_\mathrm{PEO}/V)\,p_\mathrm{EO}^\mathrm{free}\,C_\mathrm{R}$ versus the observed $\tau_3$, we find a 
nearly perfect linear relationship (correlation coefficient $-0.99$) within the statistical error of $\tau_3$. 
This demonstrates that $\tau_3$ is essentially determined by three factors: the volume fraction of PEO and the fraction of free PEO segments 
(or, alternatively, the volume concentration of free PEO segments) on one hand, and the polymer dynamics on the other hand. 

Table~\ref{tab:tau3} also shows the probability $p_\mathrm{trans,IL}$ that a TFSI-supported transfer takes place 
($p_\mathrm{trans,IL}=N_\mathrm{trans,IL}/N_\mathrm{trans}$; \ie the number of transfers in which the \Li\ is intermediately 
coordinated to TFSI only relative to the total number of transfers). 
For high IL concentrations, it becomes more likely for a lithium ion to migrate through the IL-rich regions of the electrolyte. 
Since this scenario is less likely for \Px, one can conclude that the transfer via TFSI in \Pminusx\ is mainly caused by the 
crowded PEO chains and the lower fraction of free EOs. 
Also shown in Table~\ref{tab:tau3} is the mean time $\langle\tau_\mathrm{Li}\rangle_\mathrm{IL}$ the \Li\ is coordinated to TFSI only 
in a PEO-TFSI-PEO transfer, and the average squared distance $\langle\Delta{\bf R}^2_\mathrm{Li}\rangle_\mathrm{IL}$ the lithium 
ion covers during this period. 
The values for $\langle\tau_\mathrm{Li}\rangle_\mathrm{IL}$ and $\langle\Delta{\bf R}^2_\mathrm{Li}\rangle_\mathrm{IL}$ show 
similar trends as $p_\mathrm{trans,IL}$: With increasing IL fraction and decreasing number of free EOs (\Pminusx\ electrolytes), 
both $\langle\tau_\mathrm{Li}\rangle_\mathrm{IL}$ and $\langle\Delta{\bf R}^2_\mathrm{Li}\rangle_\mathrm{IL}$ show a significant increase. 
From the Einstein relation $D_\mathrm{Li,IL}=\langle\Delta{\bf R}^2_\mathrm{Li}\rangle_\mathrm{IL}/(6\,\langle\tau_\mathrm{Li}\rangle_\mathrm{IL})$, 
we find values close to the lithium diffusion coefficient in \Pzero\ ($D_\mathrm{Li,IL}=14.014\text{~\AA$^2$}\text{ns$^{-1}$}$). 
Thus, for \Pminusx, a change of the entire lithium transport mechanism becomes visible, and the PEO-TFSI-PEO transitions may 
substantially contribute to the overall lithium MSD for sufficiently large values of $p_\mathrm{trans,IL}$ and $\langle\Delta{\bf R}^2_\mathrm{Li}\rangle_\mathrm{IL}$. 
On the other hand, from the probability $p_\mathrm{IL}$ to find a lithium ion coordinated to TFSI only at \emph{any} time (Table~\ref{tab:tau3}), 
one can conclude that the impact of the IL-mediated transport is negligible for most investigated electrolytes (see discussion in Section~\ref{sec:cmp_exp}).

\subsection{Diffusion along the Chain}
\label{sec:diff_chain}

\begin{figure}
 \centering
 \includegraphics[scale=0.3]{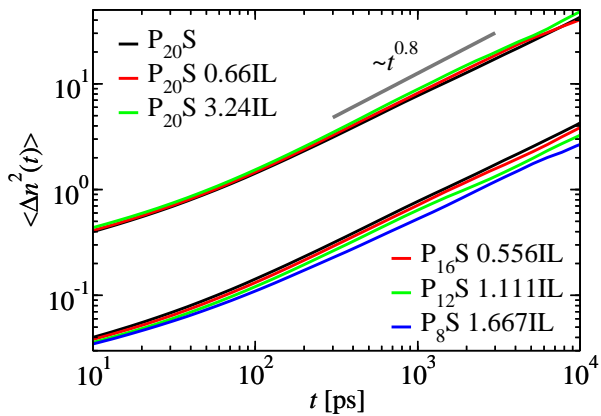}
 \caption{Mean square change of the average EO index $\langle\Delta n^2(t)\rangle$ for 
          the \Px\ and \Pminusx\ electrolytes. 
          The curves of the \Pminusx\ electrolytes have been shifted by one order of magnitude. }
 \label{fig:delta_n_msd}
\end{figure}

In order to quantify the quasi-one-dimensional random walk of the lithium ions along the PEO chains, 
all EOs have been numbered consecutively, and the mean squared change $\langle\Delta n^2(t)\rangle$ 
of the average EO index $n$ has been computed (Figure~\ref{fig:delta_n_msd}). 

Starting from about $100\text{~ps}$, the dynamics crosses over to a regime that is only slightly 
subdiffusive (\ie $\langle\Delta n^2(t)\rangle\propto t^{0.8}$). 
Qualitatively, this behavior is found for all compositions. 
For \Px, no significant change of $\langle\Delta n^2(t)\rangle$ can be observed when 
varying the IL concentration, indicating that the motion along the chain is basically 
determined by the characteristic motion of the PEO backbone, and that the surrounding 
molecules (\ie PEO chains or IL molecules) have virtually no influence on this mechanism. 
Interestingly, the magnitude of $\langle\Delta n^2(t)\rangle$ is essentially the same for 
lithium ions bound to one or to two PEO chains (not shown). 

Switching to \Pminusx, one observes that the motion along the chain becomes slower with 
increasing IL concentration, especially for \Ptwel\ and \Pei\ (Figure~\ref{fig:delta_n_msd}), 
which can be attributed to the lower fraction of free EOs in the system. 
Interestingly, the value for the exponent of $\langle\Delta n^2(t)\rangle$ remains basically 
the same in this case, and only the magnitude of the functional form $\langle\Delta n^2(t)\rangle\propto t^{0.8}$ 
becomes slightly lower. 

In principle, this type of motion could be quantified by the respective diffusion coefficient 
$D_1$ extracted from Figure~\ref{fig:delta_n_msd} according to the Einstein relation 
\begin{equation}
  D_1=\lim_{t\rightarrow\infty}\frac{\langle\Delta n^2(t)\rangle}{2\,t}\,\text{.}
\end{equation}
Unfortunately, the $\langle\Delta n^2(t)\rangle$-curves in Figure~\ref{fig:delta_n_msd} are not 
purely diffusive, whereas on larger time scales the statistics are too bad, since the ions 
are additionally transferred between distinct PEO chains, thus making it impossible to track 
the motion along the chain any further. 
In order to estimate the net effect of this mechanism, \ie the number of traversed monomers 
during $\tau_3$, $D_1$ was approximately estimated as 
\begin{equation}
 \label{eq:D1}
 D_1(\tau_3)=\frac{\langle\Delta n^2(\tau_3)\rangle}{2\,\tau_3}\,\text{.}
\end{equation}
Since the maximum observation time in Figure~\ref{fig:delta_n_msd} is lower than the $\tau_3$-values, 
we assumed that the scaling $\langle\Delta n^2(t)\rangle\propto t^{0.8}$ persists until $t=\tau_3$, 
and simply extrapolated the curves in Figure~\ref{fig:delta_n_msd}. 
Subsequently, the time scale $\tau_1$ was calculated by the respective expression from the 
transport model~\cite{MaitraHeuerPRL2007}: 
\begin{equation}
 \tau_1=\frac{\left(N-1\right)^2}{\pi^2\,D_1}
\end{equation}

\begin{table*}
 \centering
 \small
 \begin{tabular}{l c c c c c c}
  \hline
  \hline
  system & $\tau_1$ [ns] & $\langle{\bf R}_\mathrm{e}^2\rangle_\mathrm{eff}$ [$\text{\AA$^2$}$] & $\tau_\mathrm{R}$ [ns] & $\tau_2$ [ns] & $C_\mathrm{R}$ [$\text{\AA$^2$ ns$^{-1/2}$}$] & $C_2$ [$\text{\AA$^2$ ns$^{-1/2}$}$] \\
  \hline
  PEO & - & $1979$ & $22$ & - & $151.5$ & - \\
  \hline
  \Ptwen\ & $147$ & $1662$ & $45$ & $167$ & $89.0$ & $46.2$ \\
  \hline
  \multicolumn{7}{c}{ \Px\ } \\
  \hline
  \Psome\ & $140$ & $1570$ & $37$ & $89$ & $92.7$ & $59.8$ \\
  \Pmore\ & $127$ & $1571$ & $24$ & $68$ & $115.2$ & $68.4$ \\
  \hline
  \multicolumn{7}{c}{ \Pminusx\ } \\
  \hline
  \Psixt\ & $181$ & $1479$ & $43$ & $145$ & $81.0$ & $44.1$ \\
  \Ptwel\ & $208$ & $1359$ & $35$ & $145$ & $82.5$ & $40.5$ \\
  \Pei\ & $301$ & $1151$ & $28$ & $104$ & $78.1$ & $40.5$ \\
  \hline
  \hline
 \end{tabular}
 \normalsize
 \caption{Parameters characterizing the two intramolecular transport mechanisms (see text for further explanation). 
          The values for pure PEO are also shown for comparison. }
 \label{tab:tau12}
\end{table*}

Table~\ref{tab:tau12} summarizes the resulting $\tau_1$-values. 
For the \Pminusx\ electrolytes, one clearly observes an increase of $\tau_1$ with 
increasing IL concentration, since more and more EOs become involved in complexes with 
lithium, and the PEO backbone becomes less mobile. 
Contrarily, for the \Px\ electrolytes, $\tau_1$ decreases only slightly, which can mainly be 
attributed to the dependence of $\tau_1$ on $\tau_3$ (Equation~\ref{eq:D1}). 
In total, the motion along the chain basically remains unaltered in these electrolytes, 
as also suggested by Figure~\ref{fig:delta_n_msd}.

\section{Polymer Dynamics}
\label{sec:polymer_dynamics}

\subsection{Segmental Dynamics}
\label{sec:segmental_dynamcis}

\begin{figure*}
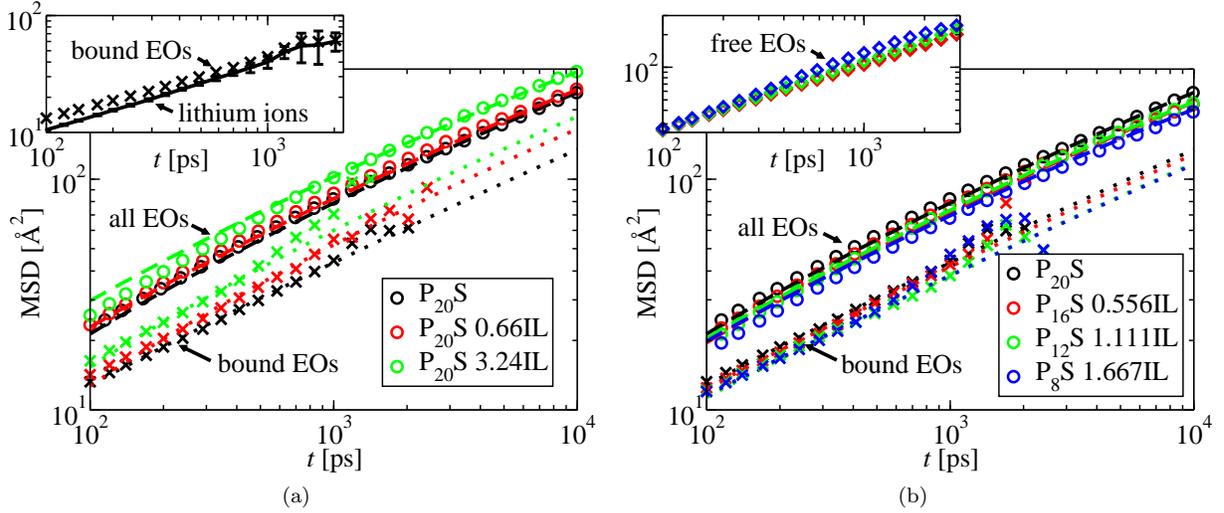

 \centering
 \subfigure[ \label{fig:EO_Li_msd_x} ]{ \includegraphics[scale=0.3]{EO_Li_msd_add} }
 \subfigure[ \label{fig:EO_Li_msd_minusx} ]{ \includegraphics[scale=0.3]{EO_Li_msd_sub} }
 \caption{MSDs of the average EOs (circles), bound EOs (diamonds), lithium ions 
          bound to these EOs ($\left|\Delta n\left(t\right)\right|\leq 1$, solid lines in the inset of Figure~\ref{fig:EO_Li_msd_x}). 
          The inset of Figure~\ref{fig:EO_Li_msd_minusx} shows the MSD of the central EOs of a PEO segment consisting of at least seven free EOs. 
          All MSDs have been computed in the center-of-mass frame of the PEO chain. 
          The dashed and the dotted lines show the respective Rouse fits. }
 \label{fig:EO_Li_msd}
\end{figure*}

In order to quantify the segmental polymer motion, the MSDs $\langle\Delta{\bf R}^2(t)\rangle$ 
of the EOs relative to the center of mass of the PEO chain have been calculated 
(Figure~\ref{fig:EO_Li_msd}). 
This quantity has been computed for all EOs (\ie irrespective of the presence of an ion), 
for EOs bound to a lithium ion and for the respective attached ions. 
The criterion to consider a cation or EO as bound was that the average EO 
index of the ion did not change more than one, \ie $\left|\Delta n(t)\right|\leq 1$ 
for all time frames during $t$. 
For the bound EOs, no further distinction between additional coordinations 
of the lithium ion to \eg another PEO chain or a TFSI molecule was made. 
Thus, these effects are already implicitly contained in the curves in Figure~\ref{fig:EO_Li_msd}. 
Of course, one might wonder if cations bound to two PEO molecules show the same dynamical characteristics as 
ions bound to single polymer chain, as the $\text{Li}(\text{PEO})_2$ complexes could be viewed as temporary crosslinks. 
A more detailed analysis (published in the appendix of our previous study~\cite{DiddensMacroLett2013}) indeed revealed that 
there is a conceptual difference between these two coordinations. 
However, the temporary crosslinks do not affect the polymer dynamics, but rather the motion along the chain: 
Whereas lithium ions coordinated to one chain are transported by both intramolecular mechanisms, 
ions bound to two chains experience the polymer dynamics only. 
This is due to the fact that the PEO chains move reptation-like along their own contour past the cation, which results in 
a non-zero $\langle\Delta n^2(t)\rangle$ (Figure~\ref{fig:delta_n_msd}), but does not contribute to the overall lithium transport. 
Since there is no qualitative effect of the coordination on the polymer dynamics (possibly also due to the fact that the 
exchange between both coordination types is sufficiently fast), we proceed to characterize the average dynamics of the bound 
PEO segments with a single $\tau_2$ only. 

The average EOs (circles) show typical Rouse-like motion with the 
characteristic relaxation time $\tau_\mathrm{R}$. 
The dynamics of the bound EOs (diamonds) is qualitatively the same 
but protracted. 
Therefore, it is possible to characterize the intramolecular dynamics 
of the bound EOs by using a larger, effective Rouse time $\tau_2$. 
The lithium ions attached to these EOs (inset, solid lines) closely 
follow the bound EOs, which gives evidence for their cooperative motion. 
On short time scales, the MSD of the EOs is larger than the lithium MSD 
due to the additional internal degrees of freedom of the PEO backbone, 
but the MSD of the bound cations catches up on longer time scales 
(approximately $1\text{~ns}$). 
Due to the collective motion, $\tau_2$ also characterizes the dynamics of the attached lithium ions. 

Figure~\ref{fig:EO_Li_msd} also shows the Rouse fits 
\begin{equation}
 \label{eq:seg_msd}
 \langle\Delta{\bf R}^2(t)\rangle = \frac{2\langle{\bf R}_\mathrm{e}^2\rangle}{\pi^2}\sum_{p=1}^\infty \frac{\left[1-\exp{\left(-\frac{tp^2}{\tau_\mathrm{R}}\right)}\right]}{p^2}
\end{equation}
for the average (dashed lines) and the bound EOs (dotted lines). 
It is important to mention that the precise value of $\tau_2$ obtained from these 
fits also depends on the value of $\langle{\bf R}_\mathrm{e}^2\rangle$ entering 
the prefactor of Equation~\ref{eq:seg_msd}. 
In order to obtain a fit consistent with the plateau value at large $t$, the MSDs 
of the average EOs were fitted using two parameters, \ie $\tau_\mathrm{R}$ and 
$\langle{\bf R}_\mathrm{e}^2\rangle_\mathrm{eff}$. 
In this way, the $\langle{\bf R}_\mathrm{e}^2\rangle$-values in Table~\ref{tab:structure} are replaced by an effective 
mean squared end-to-end vector $\langle{\bf R}_\mathrm{e}^2\rangle_\mathrm{eff}$, 
characterizing the maximum accessible intramolecular distance for the cations. 
Once $\langle{\bf R}_\mathrm{e}^2\rangle_\mathrm{eff}$ was determined from the MSDs 
of the average EOs, the MSDs of the bound EOs were fitted using this value 
in combination with a single fit parameter $\tau_2$ only. 
The fitting parameters $\langle{\bf R}_\mathrm{e}^2\rangle_\mathrm{eff}$, $\tau_\mathrm{R}$ 
and $\tau_2$ are summarized in Table~\ref{tab:tau12} (deviations from our previous study~\cite{DiddensMamol2010} on 
\Ptwen\ arise from the shorter simulation length of about $27\text{~ns}$ and the slightly modified fitting procedure). 

In case of \Px, $\langle{\bf R}_\mathrm{e}^2\rangle_\mathrm{eff}$ decreases 
only slightly when IL is added. 
The $\langle{\bf R}_\mathrm{e}^2\rangle_\mathrm{eff}$-values are in reasonable agreement with the respective 
$\langle{\bf R}_\mathrm{e}^2\rangle$-values in Table~\ref{tab:structure}. 
Compared to pure PEO, the $\langle{\bf R}_\mathrm{e}^2\rangle_\mathrm{eff}$ 
are about $16-20\text{~\%}$ lower in these systems and show good agreement 
with the respective values in Table~\ref{tab:structure}. 
This demonstrates that the PEO chains in these electrolytes behave relatively ideal, 
thus additionally validating our analysis in terms of the Rouse model. 
Both $\tau_\mathrm{R}$ and $\tau_2$ decrease significantly, clearly indicating that 
the dynamics of the PEO segments becomes faster with increasing IL concentration. 
Therefore, the IL can be regarded as plasticizer. 
Due to the enhanced polymer mobility, the lithium ions also move faster while they 
are coordinated to a specific chain, and the overall lithium MSD increases as observed 
in Figure~\ref{fig:msd_Li}. 
A similar observation has been made experimentally for other plasticizers like 
ethylene/propylene carbonate~\cite{BandaraElectrochimActa1998,KimSSI2002} or short PEO 
chains embedded in a high-molecular weight matrix~\cite{BorghiniElectrochimActa1996,KimSSI2002}.
Naturally, this effect has also been observed for ILs in the 
experiments~\cite{PasseriniJoost} that motivated the present work. 

For \Pminusx, $\langle{\bf R}_\mathrm{e}^2\rangle_\mathrm{eff}$ clearly decreases when the PEO 
chains are successively substituted by IL molecules. 
The mismatch with the $\langle{\bf R}_\mathrm{e}^2\rangle$-values in Table~\ref{tab:structure} results 
from the different averaging procedure: 
Whereas $\langle{\bf R}_\mathrm{e}^2\rangle$ is determined by the outer monomers only and thus 
rather characterizes the global polymer structure, $\langle{\bf R}_\mathrm{e}^2\rangle_\mathrm{eff}$ 
contains contributions from all monomers, making it more susceptible to structural features on short 
and intermediate length scales. 
Note that the latter quantity is also statistically much better defined, since the average is taken over all 
EOs instead of the outermost EOs only. 
The decrease of $\langle{\bf R}_\mathrm{e}^2\rangle_\mathrm{eff}$ can be rationalized by the 
fact that more and more lithium ions coordinate to a specific PEO chain, thus diminishing the 
\emph{average} intramolecular distances, and, consequently, the maximum accessible displacement 
for most ions. 
At the same time, it is still possible that the PEO chains are stretched on a global scale, which might 
arise from \eg the electrostatic repulsion between distinct ions at a given chain (although for \Pminusx, 
one cannot make clear predictions within the uncertainties of $\langle{\bf R}_\mathrm{e}^2\rangle$ in 
Table~\ref{tab:structure}). 
In combination with the decline of $\langle{\bf R}_\mathrm{e}^2\rangle_\mathrm{eff}$, 
the values for $\tau_\mathrm{R}$ and $\tau_2$ also decrease when the IL concentration 
is raised (Table~\ref{tab:tau12}). 
However, it must be pointed out that the $\tau_\mathrm{R}$- and the $\tau_2$-values 
of different electrolytes can only be quantitatively compared in conjunction with 
the precise value of $\langle{\bf R}_\mathrm{e}^2\rangle_\mathrm{eff}$. 
Therefore, a direct comparison of $\tau_\mathrm{R}$ and $\tau_2$ is only valid for similar 
$\langle{\bf R}_\mathrm{e}^2\rangle_\mathrm{eff}$, as in the case of \Px. 
This is also highlighted by the observations from Figure~\ref{fig:EO_Li_msd_minusx}, where 
the EOs (both types) rather become slower with the IL fraction. 

It is convenient to express $\langle{\bf R}_\mathrm{e}^2\rangle_\mathrm{eff}$ and $\tau_\mathrm{R}$ 
(or $\tau_2$) within a single meaningful number. 
To this purpose, the prefactor of the characteristic Rouse-like motion for intermediate 
time scales (\ie $\tau_\mathrm{R}/N^2\ll t \ll \tau_\mathrm{R}$), 
$\langle\Delta{\bf R}^2(t)\rangle=(2\langle{\bf R}_\mathrm{e}^2\rangle/\pi^{3/2})\sqrt{t/\tau_\mathrm{R}}$, 
can be calculated, making it possible to directly measure the magnitude of the MSDs. 
Table~\ref{tab:tau12} displays the constants $C_\mathrm{R}=2\langle{\bf R}_\mathrm{e}^2\rangle_\mathrm{eff}/(\pi^3 \tau_\mathrm{R})^{1/2}$ 
and $C_2=2\langle{\bf R}_\mathrm{e}^2\rangle_\mathrm{eff}/(\pi^3 \tau_2)^{1/2}$ for the individual systems, 
thereby quantifying the mobilities of the average and the bound EOs, respectively. 
Here, one indeed observes the same trends as from the direct comparison of the MSDs: 
For \Px, the segmental mobility expressed by $C_\mathrm{R}$ and $C_2$ clearly increases 
with the IL concentration, whereas it slightly decreases in case of the \Pminusx\ 
electrolytes, meaning that also the lithium dynamics decreases. 
This behavior can again be explained by the reduced flexibility of the PEO backbone, 
since the fraction of bound EOs increases when the PEO chains are gradually substituted by IL. 
Nevertheless, the plasticizing is also present for \Pminusx. 
This can be seen from the inset of Figure~\ref{fig:EO_Li_msd_minusx}, which shows the MSDs of the central EOs 
of all those PEO segments consisting of at least seven free EOs. 
Again, one observes an increase in the polymer dynamics with increasing IL fraction. 
However, since the fraction of free EOs and the probability to encounter an entirely free PEO segment 
(cf. $p_\mathrm{EO}^\mathrm{free}$ in Table~\ref{tab:tau3}) is significantly lower, the plasticizing is barely 
visible in \Pminusx. 
Rather, for the average EOs, the progressive coordination of lithium ions to the PEO chains screens the 
plasticizing effect of the IL. 

Finally, it is worth mentioning that for both investigated types of electrolytes, the center-of-mass dynamics of the PEO 
chains becomes faster with increasing IL concentration. 
(The impact of this increase on the lithium diffusion coefficient will be discussed in Section~\ref{sec:cmp_exp}). 
However, this observation can be easily rationalized by the fact that motion of the individual chains is less hindered by 
other chains when the PEO concentration is reduced.

\subsection{Hydrodynamic Interactions}

\begin{figure}
  \centering
  \includegraphics[scale=0.3]{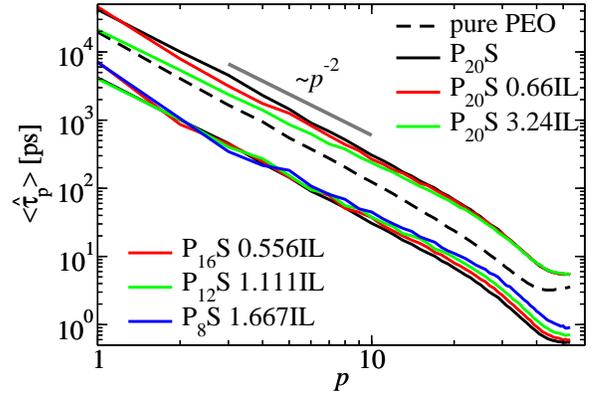}
  \caption{Average Rouse mode relaxation times $\langle\hat{\tau}_p\rangle$ in dependence of the mode number $p$. 
          The curves of the \Pminusx\ electrolytes have been shifted by one order of magnitude. }
  \label{fig:taup}
\end{figure}

An especially interesting effect of the IL on the polymer dynamics can be observed 
from the scaling of the Rouse mode relaxation times. 
The correlators $\langle{\bf X}_p(0){\bf X}_p(t)\rangle$ display a stretched exponential 
relaxation (not shown), and a Kohlrausch-Williams-Watts fit, 
$\langle{\bf X}_p(0){\bf X}_p(t)\rangle/\langle{\bf X}_p^2\rangle = \exp(-({t}/{\hat{\tau}_p})^\beta)$, 
was used to extract the parameters $\hat{\tau}_p$ and $\beta$ 
(here, the hats have been used to avoid confusion with the time scales $\tau_1$, $\tau_2$ and $\tau_3$ 
of the lithium ion transport model). 
The mean relaxation time $\langle\hat{\tau}_p\rangle$ is then given by 
$\langle\hat{\tau}_p\rangle=\hat{\tau}_p\beta^{-1}\Gamma(\beta^{-1})$, where $\Gamma$ 
is the gamma function. 
The resulting scaling of $\langle\hat{\tau}_p\rangle$ is shown in Figure~\ref{fig:taup}. 
Deviations from the values for $\tau_\mathrm{R}\equiv \langle\hat{\tau}_1\rangle$ in Table~\ref{tab:tau12} 
result from the fact that the segmental MSD reflects an average over all modes, whereas 
$\langle\hat{\tau}_1\rangle$ is determined from the first mode only. 

For low mode numbers, both pure PEO as well as \Ptwen\ exhibit the same scaling 
of approximately $\langle\hat{\tau}_p\rangle\propto p^{-2}$, thereby indicating that the 
global modes are Rousean. 
The absolute values of the $\langle\hat{\tau}_p\rangle$ are systematically higher for 
\Ptwen\ due to the decelerating effect of the lithium ions. 

In contrast to this, one observes a clear decrease of the relaxation times at low $p$ 
for \Pmore\ as compared to \Ptwen. 
As already observed from the fits of the segmental MSDs (Figure~\ref{fig:EO_Li_msd}), \Pmore\ 
approximately has the same value for $\tau_\mathrm{R}$ as pure PEO, whereas the local 
$\langle\hat{\tau}_p\rangle$ ($p\gg 1$) are basically the same as for the binary \Ptwen. 
However, a faster global relaxation combined with a similar local relaxation can only be 
realized by a less steep dependence of $\langle\hat{\tau}_p\rangle$ on $p$. 
These observations are compatible with a scenario in which hydrodynamic interactions, arising 
from the presence of the IL, contribute to the overall polymer motion. 
From an analytical point of view, these interactions can readily be taken into account via the Zimm model~\cite{ZimmJCP1956,DoiEdwards}. 
In its simplest form, \ie when no swelling of the chains is present (cf. $\langle{\bf R}_\mathrm{e}^2\rangle$ 
in Table~\ref{tab:structure} and $\langle{\bf X}_p^2\rangle$ in Figure~\ref{fig:Xp2}), the equilibrium 
distribution of the chain conformation is Gaussian as in the Rouse model. 
Therefore, the static polymer properties remain unaffected, and only the dynamics is 
altered, leading to a scaling of the relaxation times of ${\hat\tau}_p\propto p^{-3/2}$. 

Interestingly, for \Pminusx, the scaling of the $\langle\hat{\tau}_p\rangle$ for 
intermediate $p$ does not change significantly. 
However, the uncertainties of these curves increase when the PEO chains are successively 
replaced by IL (up to $60\text{~\%}$ in the worst case, \ie $p=1$ for \Pei). 
For the local modes, the $\langle\hat{\tau}_p\rangle$ increase with the IL concentration, again 
highlighting the reduced flexibility of the PEO backbone. 
Therefore, in total, the replacement of PEO by IL rather leads to a decrease of the polymer 
dynamics, and the plasticizing effect of the IL is not sufficient to compensate for the 
slowing down of the PEO segments caused by the attached lithium ions. 

Although the scaling of the $\langle\hat{\tau}_p\rangle$ is rather Zimm-like~\cite{ZimmJCP1956} 
(\ie $\langle\hat{\tau}_p\rangle\propto p^{-3/2}$) for \Px, especially for \Pmore, the segmental 
MSD nevertheless shows the characteristic Rousean proportionality $\text{MSD}\propto t^{1/2}$ 
for $t\geq 1\text{~ns}$ (Figure~\ref{fig:EO_Li_msd}) instead of the respective Zimm scaling~\cite{DoiEdwards} 
$\text{MSD}\propto t^{2/3}$. 
Only on time scales up to about $1\text{~ns}$, the MSD of the average EOs is lower and slightly steeper 
than the Rouse predictions for \Pmore. 
This is in clear contrast to \Ptwen, for which the PEO curve is larger than the Rouse prediction 
due to the additional internal degrees of freedom of the PEO backbone. 
Thus, this feature reflects another minor imprint of the hydrodynamic interactions. 
On larger length scales however, the hydrodynamic interactions are screened, since the monomer 
concentration in the PEO-IL solutions is still relatively high. 
Consequently, the chains interpenetrate strongly as in the pure polymer melt, and the momentum 
transfer via the low-molecular solvent acts only on short length scales. 
A similar dynamical behavior has also been reported previously for semidilute solutions of model 
polymers~\cite{DuenwegPRE2001} simulated by a combination of stochastic Langevin dynamics and 
Lattice Boltzmann techniques~\cite{DuenwegJCP1999}. 
Since the motion on larger scales is still Rousean, we also left Equation~\ref{eq:seg_msd} unchanged in 
order to extract $\tau_2$ for the \Px\ electrolytes.

\section{Comparison with Experiments}
\label{sec:cmp_exp}

Finally, we compare the predictions of the lithium transport model to the experimental data 
taken from ref.~\citenum{PasseriniJoost}. 
Of course, a direct comparison of $D_\mathrm{Li}$ as extracted from Figure~\ref{fig:msd_Li} is impossible, 
since the center-of-mass motion of the short PEO chains ($N=54$) in our simulations significantly contributes to 
the lithium diffusion, whereas most experimental studies focus on the long-chain limit 
(\eg $M_\mathrm{w}=4.000.000\text{~g/mol}$ in ref.~\citenum{PasseriniJoost}). 
However, due the explicit $N$-dependence of the three time scales~\cite{MaitraHeuerPRL2007}, 
\ie $\tau_1\propto N^2$, $\tau_2\propto N^2$ and $\tau_3\propto N^0$, and using the fact that the 
center-of-mass motion vanishes for $N\rightarrow\infty$, this problem can be easily circumvented. 

For the scaling of $\tau_2$, entanglement effects may become relevant, which would manifest 
itself by a steeper dependence of the polymer relaxation time on $N$ for chain lengths larger than 
the entanglement length $N_\mathrm{e}$. 
However, if $\tau_3<\tau_\mathrm{e}$, meaning that the lithium ion leaves the chain before the 
latter begins to reptate, the dynamical contribution to the lithium ions is still Rousean~\cite{DoiEdwards}. 
In the case of PEO, experiments revealed that the entanglement regime sets in from about 
$N\approx 75$~\cite{AppelMamol1993,ShiSSI1993}. 
Based on these observations, one can estimate $\tau_\mathrm{e}$ according to 
$\tau_\mathrm{e}=\tau_\mathrm{R}(N=75)=\tau_\mathrm{R}(75/54)^2$. 
For \Ptwen, this leads to $\tau_\mathrm{e}\approx 87\text{~ns}$, which is substantially 
larger than $\tau_3$. 
Also in case of the highly plasticized \Pmore\ one finds $\tau_\mathrm{e}\approx 46\text{~ns}>\tau_3$. 
Therefore, the lithium ion leaves the PEO chain before the tube constraints become noticeable, 
and the transport model can also be applied in the limit of long chains. 

Due to the renewal events (\ie the interchain transfer), $D_\mathrm{Li}$ can be expressed as 
\begin{equation}
 \label{eq:D_renewal}
 D_\mathrm{Li}(N)=\frac{\langle g_{12}(N,\tau_3)\rangle_\mathrm{M3}}{6\tau_3}+D_\mathrm{PEO}(N)\,\text{,}
\end{equation}
where $\langle g_{12}(\tau_3)\rangle_\mathrm{M3}$ corresponds to an averaged, elementary 
step length between two successive renewal events (abbreviated as M3), and is characterized 
by the intramolecular transport mechanisms (\ie the diffusion along the chain and the polymer dynamics). 
The center-of-mass motion of the PEO chains with length $N$ is characterized by the diffusion 
coefficient $D_\mathrm{PEO}$. 
As shown in earlier work~\cite{MaitraHeuerPRL2007}, $g_{12}$ can be described by a Rouse-like expression 
\begin{equation}
 \label{eq:g12model}
 g_{12}(t) = \frac{2\langle{\bf R}_\mathrm{e}^2\rangle}{\pi^2}\sum_{p=1}^\infty\frac{\left[1-\exp{\left(-\frac{tp^2}{\tau_{12}}\right)}\right]}{p^2}\,\text{,}
\end{equation}
where $1/\tau_{12}=1/\tau_1+1/\tau_2$ is a combined relaxation rate due to both mechanisms. 
As argued in Section~\ref{sec:polymer_dynamics}, lithium ions coordinated to two PEO chains experience 
no effective transport due to the diffusion along the chain (see also supporting information of ref.~\citenum{DiddensMacroLett2013}). 
However, with respect to Equation~\ref{eq:g12model}, this effect can be easily captured. 
For a sufficiently fast exchange between coordinations to one and to two PEO chains, the average intramolecular 
lithium MSD $\bar{g}_{12}$ is simply determined by the structural quantities $r_1$ and $r_2$ (Table~\ref{tab:structure}): 
\begin{equation}
 \label{eq:g12_g2model}
 \bar{g}_{12}(t) = r_\mathrm{1}\,g_{12}(t) + r_2\,g_{2}(t)
\end{equation}
That is, for the fraction $r_1$ of cations bound to one PEO chain, Equation~\ref{eq:g12model} remains valid, whereas for 
ions bound to two chains with $\tau_{1,\mathrm{\,2\,PEO}}\rightarrow\infty$, only $\tau_2$ is important. 
For exponentially distributed residence times at the distinct PEO chains (which is fulfilled to a good approximation 
in our simulations except for shortest residence times lower than $300\text{~ps}$, which account only for a few percent), 
one can write down an analytical expression for the average MSD at a given chain: 
\begin{equation}
 \begin{aligned}
  \langle g_{12}(\tau_3)\rangle_\mathrm{M3} &= \frac{1}{\tau_3}\int_{0}^\infty dt\exp{\left(-\frac{t}{\tau_3}\right)}\,g_{12}(t) \\
                                            &= \frac{2\langle{\bf R}_\mathrm{e}^2\rangle}{\pi^2}\sum_{p=1}^\infty\frac{1}{p^2}\left[1-\frac{1}{p^2\frac{\tau_3}{\tau_{12}}+1}\right]
 \end{aligned}
 \label{eq:g12_expdist}
\end{equation}
Solving this expression numerically, values for $D_\mathrm{Li}$ were obtained for $N=54$ ($D_\mathrm{Li}^\mathrm{sim}$) and 
$N\rightarrow\infty$ ($D_\mathrm{Li}^{\infty}$), which are summarized in Table~\ref{tab:D_Li} (here, a correction similar to 
Equation~\ref{eq:g12_g2model} has been applied). 
For the calculation of $D_\mathrm{Li}^\mathrm{sim}$, $D_\mathrm{PEO}$ was simply extracted from the simulation. 
When comparing $D_\mathrm{Li}^\mathrm{sim}$ with the lithium diffusion coefficients directly extracted from Figure~\ref{fig:msd_Li}, 
we find reasonable agreement between both values (although the diffusion coefficients from the MSDs are about $5-20\text{~\%}$ 
higher, which is due to the fact that the lithium motion has not yet become fully diffusive and the statistical uncertainties 
are relatively large in the long-time regime). 
However, when using the transport model to calculate the lithium MSDs as in our previous work~\cite{DiddensMacroLett2013} 
(which was done by simulating a random-walk motion whose step length was given by Equation~\ref{eq:g12model}, and in which random reorientations 
at given time steps sampled from a Poisson process modeled the renewal events), we find quantitative agreement within the error bars for 
all systems, showing that the predictions of our model are consistent with the numerical data. 

\begin{table*}
 \centering
 \footnotesize
 \begin{tabular}{l c c c l}
  \hline
  \hline
  system & $D_\mathrm{Li}^\mathrm{sim}$ [$\text{\AA$^2$}\text{ns}^{-1}$] & $D_\mathrm{Li}^{\infty}$ [$\text{\AA$^2$}\text{ns}^{-1}$] & \multicolumn{2}{c}{$D_\mathrm{Li}^\mathrm{exp}$ [$\text{\AA$^2$}\text{ns}^{-1}$] (ref.~\citenum{PasseriniJoost})} \\
  \hline
  \Ptwen\ & $2.945$ ($2.947$) & $1.947$ ($1.949$) & $0.052$ & (\Ptwoneone) \\
  \hline
  \multicolumn{5}{c}{ \Px\ } \\
  \hline
  \Psome\ & $3.542$ ($3.545$) & $2.309$ ($2.313$) & $0.118$ & (\Ptwonetwo) \\
  \Pmore\ & $4.257$ ($4.292$) & $2.392$ ($2.434$) & $0.126$ & (\Ptwonefour) \\
  \hline
  \multicolumn{5}{c}{ \Pminusx\ } \\
  \hline
  \Psixt\ & $2.573$ ($2.581$) & $1.511$ ($1.520$) & $0.051$ & (\Ptoneone) \\
  \Ptwel\ & $2.716$ ($2.765$) & $1.311$ ($1.366$) & $0.063$ & (\Ptonetwo) \\
  \Pei\ & $2.812$ ($3.092$) & $1.166$ ($1.487$) & $0.046$ & (\Pfoneone) \\
  \hline
  \hline
 \end{tabular}
 \normalsize
 \caption{Lithium diffusion coefficient $D_\mathrm{Li}$ calculated by the lithium transport model (in particular Equations~\ref{eq:D_renewal} 
          and \ref{eq:g12_expdist}) for $N=54$ ($D_\mathrm{Li}^\mathrm{sim}$) and $N\rightarrow\infty$ ($D_\mathrm{Li}^{\infty}$). 
          The values in parentheses for $D_\mathrm{Li}^\mathrm{sim}$ and $D_\mathrm{Li}^{\infty}$ are corrected values, for which 
          the lithium migration through IL-rich regions has been taken into account. 
          Experimental values~\cite{PasseriniJoost} for similar compositions at $T=323\text{~K}$ are also shown. }
 \label{tab:D_Li}
\end{table*}

In the following, we will compare our results with the experimental data ($D_\mathrm{Li}^\mathrm{exp}$) 
at $T=323\text{~K}$ taken from ref.~\citenum{PasseriniJoost}. 
Despite the temperature gap between simulation and experiment, we observe identical trends: 
For a similar dilution series, in which the IL is {\it added} to \Ptwen, $D_\mathrm{Li}^\mathrm{exp}$ 
also increases significantly with increasing IL content. 
In contrast to this, $D_\mathrm{Li}^\mathrm{exp}$ rather decreases when the PEO chains are {\it substituted} 
by IL, and, consequently, the $\text{EO}:\text{Li}$ ratio decreases. 

As already discussed in context with the segmental PEO motion (Section~\ref{sec:segmental_dynamcis}), these observations can be understood as follows: 
In case of \Px, the IL acts as a plasticizer, and the enhanced polymer dynamics leads to a faster lithium ion dynamics. 
On the other hand, when the $\text{EO}:\text{Li}$ ratio simultaneously decreases as for \Pminusx, the remaining PEO chains 
become more and more rigid, which approximately cancels the plasticizing effect of the IL. 

Thus, two opposing trends related to the segmental mobility can be found in ternary polymer 
electrolyte-ionic liquid mixtures: the slowdown of the polymer motion with decreasing $\text{EO}:\text{Li}$ 
ratio (also encountered in the binary systems) and the plasticizing due to the IL, leading to an 
enhancement of the polymer dynamics (\ie both the segmental and the center-of-mass motion). 
Naturally, for the technologically relevant long-chain limit, the enhancement of the center-of-mass 
motion of the PEO chains is irrelevant. 
For these reasons, one should keep in mind that especially the polymer segments remain mobile enough when 
designing new electrolyte materials. 

It should be pointed out that the $D_\mathrm{Li}$-values in Table~\ref{tab:D_Li} do not contain the contribution from those lithium ions 
which are coordinated by IL molecules only and thus move faster. 
In particular for the electrolytes \Ptwel\ and \Pei, for which this additional, fourth mechanism becomes relevant (Table~\ref{tab:tau3}), 
this would lead to noticeably higher $D_\mathrm{Li}$-values in Table~\ref{tab:D_Li}. 
Here, a correction to $D_\mathrm{Li}$ can easily be estimated: 
At a given time, the fraction of lithium ions coordinated by TFSI only is given by $p_\mathrm{IL}$ (Table~\ref{tab:tau3}). 
Since we found that dynamics of these ions is similar to that in \Pzero, one can use the respective lithium diffusion coefficient 
($D_\mathrm{Li,IL}=14.014\text{~\AA$^2$}\text{ns$^{-1}$}$) to estimate their contribution to the overall diffusion coefficient, 
whereas the relative contribution of the remaining lithium ions is still described by Equation~\ref{eq:D_renewal}: 
\begin{equation}
 \label{eq:D_renewal_corr}
 D_\mathrm{Li}^\mathrm{corr}(N) = (1-p_\mathrm{IL})\,D_\mathrm{Li}(N) + p_\mathrm{IL}\,D_\mathrm{Li,IL}
\end{equation}
The respective corrected values for $D_\mathrm{Li}^\mathrm{sim}$ and $D_\mathrm{Li}^{\infty}$ are given in parentheses in Table~\ref{tab:D_Li}. 
As already expected from the values of $p_\mathrm{IL}$, significant contributions arise only for \Ptwel\ and \Pei\ 
($p_\mathrm{IL}=0.4\text{~\%}$ and $2.5\text{~\%}$, respectively), whereas $D_\mathrm{Li}$ for all other electrolytes remains 
basically unaltered. 

Thus, for most electrolytes, the value of $D_\mathrm{Li}$ is determined by essentially three contributions (Equation~\ref{eq:D_renewal}): 
the intrachain distance $\langle g_{12}(\tau_3)\rangle_\mathrm{M3}$ the ion covers while 
connected to the same chain, the center-of-mass motion of the PEO molecules characterized by $D_\mathrm{PEO}$ 
and the renewal events measured by $\tau_3$, which facilitate the diffusive motion of the ions in the long-time 
limit for $N\rightarrow\infty$. 
Moreover, these trends may also oppose each other as in case of \Px. 
For these reasons, it is desirable to quantify in how far the modified intrachain transport 
and the altered renewal rates contribute to the overall change of $D_\mathrm{Li}$ relative to 
the pure polymer electrolyte \Ptwen. 

\begin{table}
 \centering
 \footnotesize
 
 \subtable[$N=54$]{
 \begin{tabular}{l r r r r}
  \hline
  \hline
  system & $\langle g_{12}(\tau_3)\rangle_\mathrm{M3}$ & $\tau_3^{-1}$ & $D_\mathrm{PEO}$ & $D_\mathrm{Li}$ \\
  \hline
  \Ptwen\ & $\pm0$ & $\pm0$ & $\pm0$ & $\pm0$ \\
  \hline
  \multicolumn{5}{c}{ \Px\ } \\
  \hline
  \Psome\ & $+22$ & $-7$ & $+30$ & $+20$ \\
  \Pmore\ & $+54$ & $-29$ & $+92$ & $+45$ \\
  \hline
  \multicolumn{5}{c}{ \Pminusx\ } \\
  \hline
  \Psixt\ & $+9$ & $-32$ & $+4$ & $-13$ \\
  \Ptwel\ & $+6$ & $-40$ & $+30$ & $-8$ \\
  \Pei\ & $+5$ & $-48$ & $+50$ & $-5$ \\
  \hline
  \hline
 \end{tabular}
 }

 \subtable[$N\rightarrow\infty$]{
 \begin{tabular}{l r r r r}
  \hline
  \hline
  system & $\langle g_{12}(\tau_3)\rangle_\mathrm{M3}$ & $\tau_3^{-1}$ & $D_\mathrm{PEO}$ & $D_\mathrm{Li}$ \\
  \hline
  \Ptwen\ & $\pm0$ & $\pm0$ & - & $\pm0$ \\
  \hline
  \multicolumn{5}{c}{ \Px\ } \\
  \hline
  \Psome\ & $+28$ & $-7$ & - & $+19$ \\
  \Pmore\ & $+73$ & $-29$ & - & $+23$ \\
  \hline
  \multicolumn{5}{c}{ \Pminusx\ } \\
  \hline
  \Psixt\ & $+14$ & $-32$ & - & $-22$ \\
  \Ptwel\ & $+12$ & $-40$ & - & $-33$ \\
  \Pei\ & $+15$ & $-48$ & - & $-40$ \\
  \hline
  \hline
 \end{tabular}
 }
 
 \normalsize
 \caption{Percental changes of the intramolecular contribution $\langle g_{12}(\tau_3)\rangle_\mathrm{M3}$, 
          the renewal rates $\tau_3^{-1}$, the center-of-mass motion $D_\mathrm{PEO}$ and the resulting 
          $D_\mathrm{Li}$-value relative to \Ptwen\ for (a)~$N=54$ and (b)~$N\rightarrow\infty$. }
 \label{tab:M12_M3_contr}
\end{table}

Table~\ref{tab:M12_M3_contr} shows the percentage by which the intramolecular 
contributions ($\langle g_{12}(\tau_3)\rangle_\mathrm{M3}$), 
the center-of-mass motion of the PEO chains ($D_\mathrm{PEO}$) 
and the renewal rates ($\tau_3^{-1}$) as well as the overall $D_\mathrm{Li}$-values 
change relative to the pure polymer electrolyte \Ptwen\ for $N=54$ and $N\rightarrow\infty$. 

For \Px\ one clearly observes that the intramolecular contribution increases 
with the IL concentration, mainly as a result of the increased segmental 
mobility. 
For finite PEO chains, the enhanced $D_\mathrm{PEO}$-value additionally increases 
the lithium motion. 
In contrast to this, the contribution of the renewal processes decreases for 
these electrolytes. 
However, the plasticizing effect overcompensates this decrease, and $D_\mathrm{Li}$ 
increases for both $N=54$ and $N\rightarrow\infty$. 

Remarkably, the relative intramolecular contribution also increases for 
\Pminusx, although the enhancement is weaker than for \Px. 
At first glance, this may contradict the findings presented in Section~\ref{sec:segmental_dynamcis}, where it 
was observed that both the diffusion along the PEO chain and the segmental 
motion become slower when IL is added to the system. 
However, this mismatch can easily be resolved by the following considerations:
First, the fraction of lithium ions coordinating to one PEO chain only increases 
when going from \Ptwen\ to \Pei\ (Table~\ref{tab:structure}). 
Since lithium ions coordinated to two polymer chains experience no effective transport via this mechanism 
(supporting information in ref.~\citenum{DiddensMacroLett2013}), the diffusion along the PEO backbone becomes more efficient in this way. 
The second, more important effect leading to the observations from Table~\ref{tab:M12_M3_contr} 
is the fact that $\langle g_{12}(\tau_3)\rangle_\mathrm{M3}$ is evaluated for different 
$\tau_3$ (Equation~\ref{eq:g12_expdist}). 
Thus, with increasing $\tau_3$ also the distance $\langle g_{12}(\tau_3)\rangle_\mathrm{M3}$ 
increases, although this dependence is weaker than linear, and the overall 
$D_\mathrm{Li}$ diminishes in this case. 
Consistently, calculating the individual values for $\langle g_{12}(\tau_3)\rangle_\mathrm{M3}$ 
with a constant value of $\tau_3=17.1\text{~ns}$ as observed for \Ptwen\ (Table~\ref{tab:tau3}), the intramolecular 
contribution indeed decreases for \Pminusx. 
The only contribution which displays a clear increase is $D_\mathrm{PEO}$. 
Thus, for $N=54$, $D_\mathrm{Li}$ diminishes only slightly, whereas a significant decrease 
can be found in case of $N\rightarrow\infty$. 

So far, we focused only on the high-temperature limit which we can address in our simulations. 
Interestingly, the relative increase of $D_\mathrm{Li}^\mathrm{exp}$ upon the addition of IL becomes much more 
pronounced in the low-temperature regime~\cite{PasseriniJoost}. 
Although simulations at low temperature would be too costly, we are nevertheless able to make some statements 
about this regime. 
For example, in case of the addition of IL, it was reported that the glass-transition temperature $T_\mathrm{g}$ 
significantly decreases (up to $35\text{~K}$) with increasing IL concentration~\cite{PasseriniJoost}, which 
is a first indication that also at ambient temperatures the enhanced polymer dynamics contributes to a larger 
$D_\mathrm{Li}$. 
In contrast to this, the experimentally observed decrease of $T_\mathrm{g}$ is significantly weaker for lower 
$\text{EO}:\text{Li}$ ratios~\cite{PasseriniJoost} (which resembles the case of IL substitution). 
Simultaneously, $D_\mathrm{Li}^\mathrm{exp}$ approximately remains constant at $323\text{~K}$. 
A detailed analysis of the low-temperature regime is already underway.

\section{Conclusion}
\label{sec:conclusion}

In this work, we presented an exhaustive MD simulation study on ternary polymer electrolytes consisting of 
PEO/LiTFSI and the ionic liquid PYR$_\mathrm{13}$TFSI. 
In total, we investigated two different concentration series, namely (a)~the \emph{addition} of the ionic liquid 
to \Ptwenty, and (b)~the \emph{substitution} of the PEO chains in \Ptwenty\ by the ionic liquid, subject to the 
constraint of a constant lithium volume concentration. 
The latter electrolytes not only reflected an interpolation between \Ptwenty\ and \Pzero, but also served as 
model systems for electrolytes with lower $\text{EO}:\text{Li}$ ratios. 
In addition to the simulations, we also utilized a lithium ion transport model~\cite{MaitraHeuerPRL2007} 
to express the effect of the ionic liquid on the lithium diffusion via a few microscopic parameters. 
Based on the combined insights from the simulations and the transport model, we were able to formulate general 
statements in how far the lithium ion transport mechanism changes when the composition of the electrolyte is 
varied. 

In case of sufficiently high $\text{EO}:\text{Li}$ ratios (in particular $20:1$), we find that the lithium 
transport almost exclusively takes place at the PEO chains, and that the transport mechanism therefore is 
qualitatively the same as in binary \Ptwenty. 
For lower ratios, however, more and more lithium ions coordinate to a specific PEO chain, with the result 
that a progressive coordination by TFSI anions (both partially and totally) becomes visible. 
Accordingly, a completely new, fourth transport mechanism, in which the lithium ions are decoupled from the 
PEO chains, emerges due to the lack of free ether oxygens capable to bind the lithium ions. 
This mechanism may significantly contribute to the macroscopic lithium diffusion or even dominate it. 
From a theoretical perspective, we incorporated this mechanism into our transport model. 
For all electrolytes, we find reasonable agreement of the model predictions with the experimentally observed 
diffusion coefficients~\cite{PasseriniJoost}. 

Apart from these qualitative differences, the lithium transport is also altered on a quantitative level. 
For instance, the renewal rate $\tau_3^{-1}$, characterizing the transfer between distinct PEO chains, 
decreases with increasing ionic-liquid concentration and with decreasing fraction of free ether oxygens. 
Consequently, $\tau_3$ is especially large for diluted electrolytes with crowded PEO chains (\ie the 
\PminusxIL\ systems). 

Concerning the polymer dynamics, it turned out that the increase of the lithium diffusivity in \PxIL\ can 
be attributed to the enhanced segmental mobility of the PEO chains, and that the ionic liquid thus serves 
as a plasticizer. 
As a result of the faster PEO motion, also the lithium ions coordinating to the PEO chains become faster. 
Interestingly, the plasticization is also visible in the scaling of the Rouse-mode relaxation times, which 
exhibit a slightly Zimm-like scaling characteristic for semidilute polymer solutions. 
An opposing effect on the PEO dynamics, however, is the coordination to the lithium ions, which becomes 
especially pronounced for low $\text{EO}:\text{Li}$ ratios. 
Accordingly, the average PEO dynamics in \PminusxIL\ is slowed down by the progressive coordination of 
lithium ions, and the plasticizing effect of the ionic liquid becomes visible only for larger, entirely 
uncoordinated polymer segments. 
In summary, one therefore faces the situation that the segmental mobility of the PEO chains plays a decisive 
role in ternary polymer electrolytes: 
Whereas the lithium ions slow down the PEO chains, the ionic liquid accelerates the polymer motion. 
Thus, for the design of new battery materials, one should attempt to render the latter effect the 
dominating one.

\begin{acknowledgments}
The authors would like to thank Oleg Borodin, Nicolaas A. Stolwijk, Stefano Passerini 
and Mario Joost for helpful discussions and for providing the experimental data. 
Financial support from the NRW Graduate School of Chemistry is also greatly appreciated. 
\end{acknowledgments}

\begin{appendix}
\section{Part~A: Further Analyses}

\subsection{Simulation Box Size}

\begin{table}
 \centering
 \begin{tabular}{l c}
  \hline
  \hline
  system & $L_\mathrm{box}$ [\AA] \\
  \hline
  PEO & $40.11$ \\
  \hline
  \Ptwen\ & $35.96$ \\
  \hline
  \multicolumn{2}{c}{ \Px\ } \\
  \hline
  \Psome\ & $38.27$ \\
  \Pmore\ & $45.29$ \\
  \hline
  \multicolumn{2}{c}{ \Pminusx\ } \\
  \hline
  \Psixt\ & $35.93$ \\
  \Ptwel\ & $35.96$ \\
  \Pei\ & $36.00$ \\
  \hline
  \Pz\ & $36.33$ \\
  \hline
  \hline
 \end{tabular}
 \caption{Length of the simulation boxes for the different electrolytes. }
 \label{tab:boxlength}
\end{table}

The sizes of the individual simulation boxes are summarized in Table~\ref{tab:boxlength}. 
One clearly observes that while for the \Px\ systems the volume of the simulation box increases, 
$L_\mathrm{box}$ for the \Pminusx\ electrolytes is nearly identical for each system, meaning 
that the lithium volume concentration ($27$ lithium ions in each box) remains constant.

\subsection{Radial Distribution Functions}
\label{ssec:gofr}

\begin{figure*}
 \centering
 \subfigure[ \label{fig:gofr_LiO_PEO_x} ]{ \includegraphics[scale=0.3]{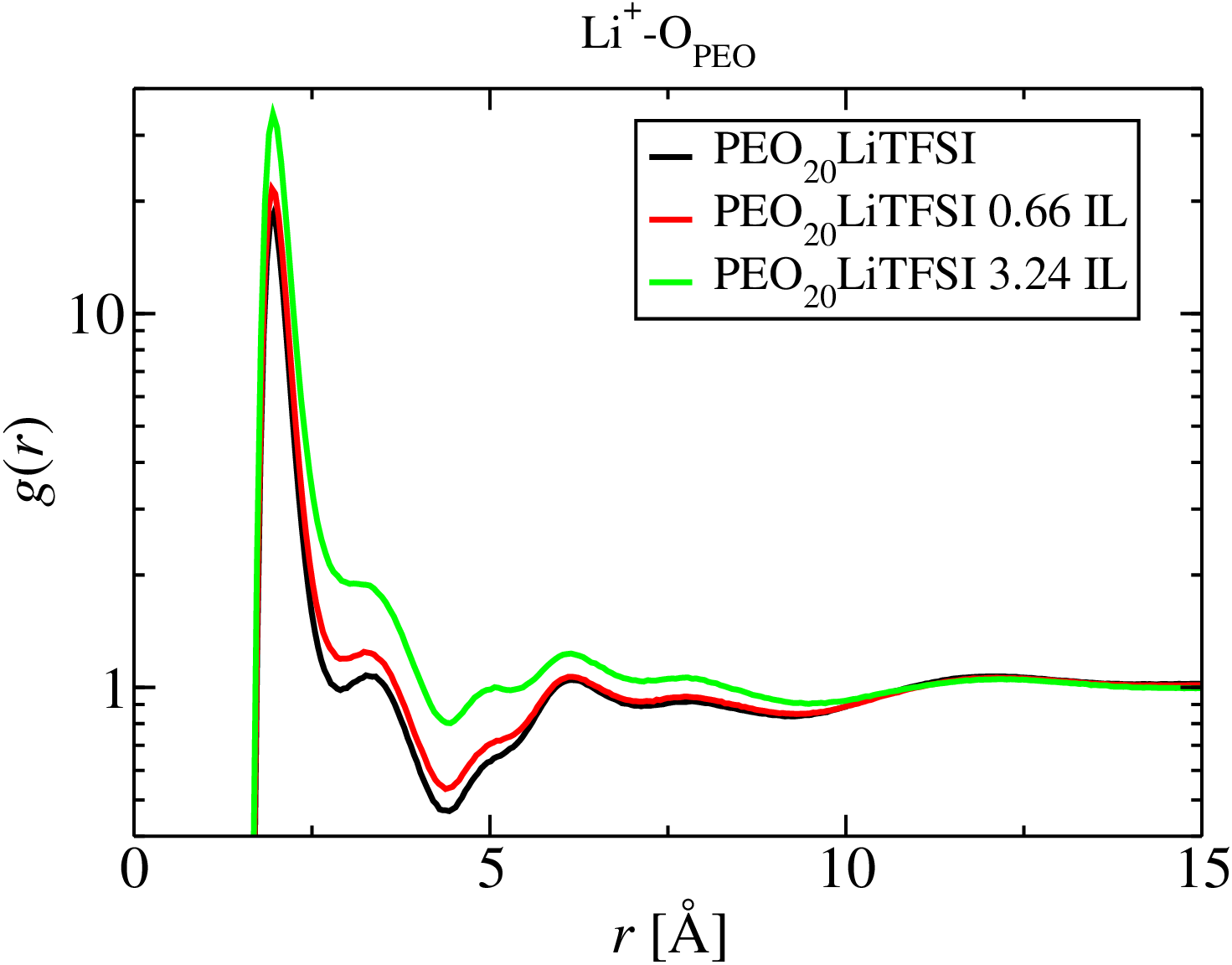} }
 \subfigure[ \label{fig:gofr_LiO_TFSI_x} ]{ \includegraphics[scale=0.3]{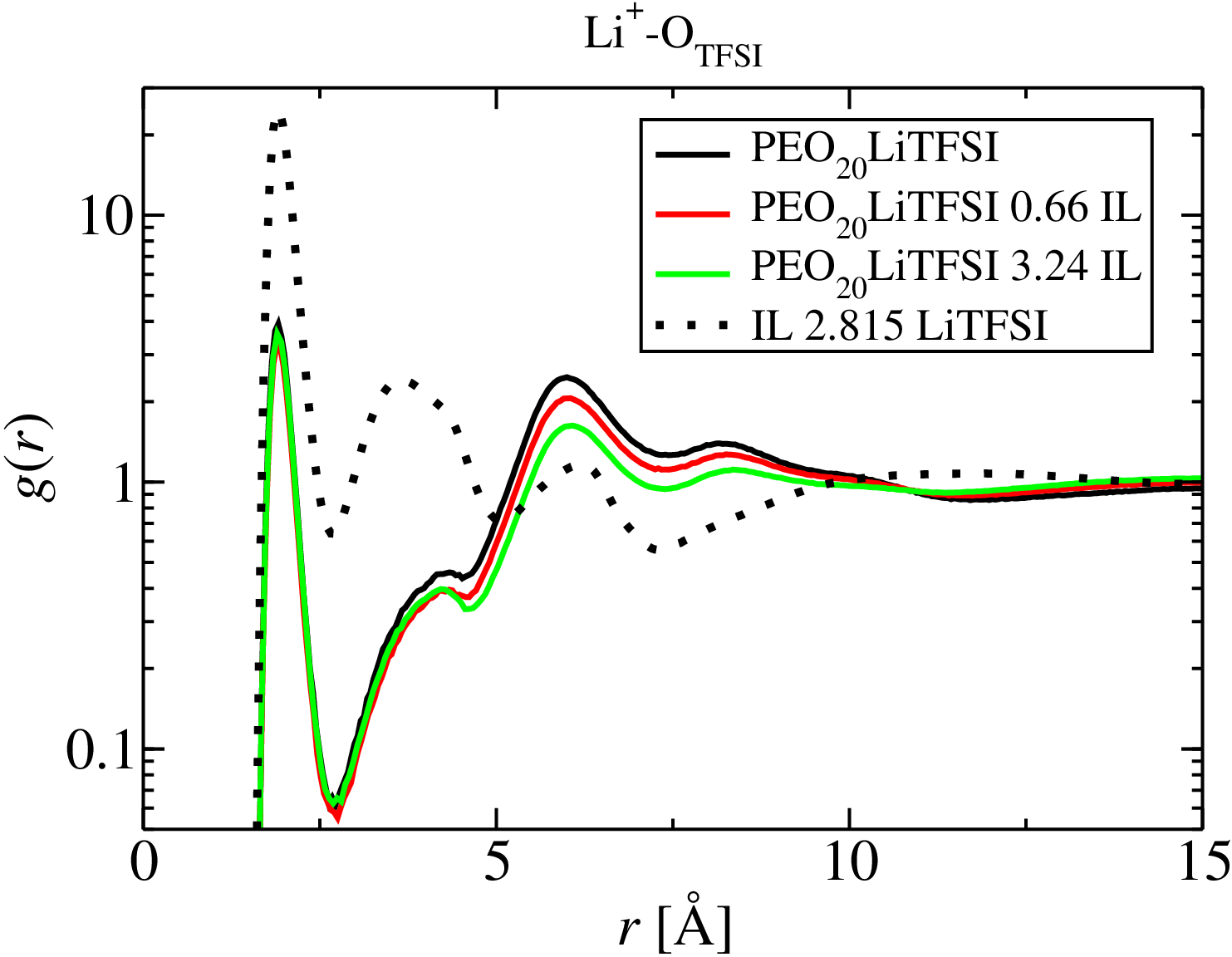} }
 \subfigure[ \label{fig:gofr_LiO_PEO_xminus} ]{ \includegraphics[scale=0.3]{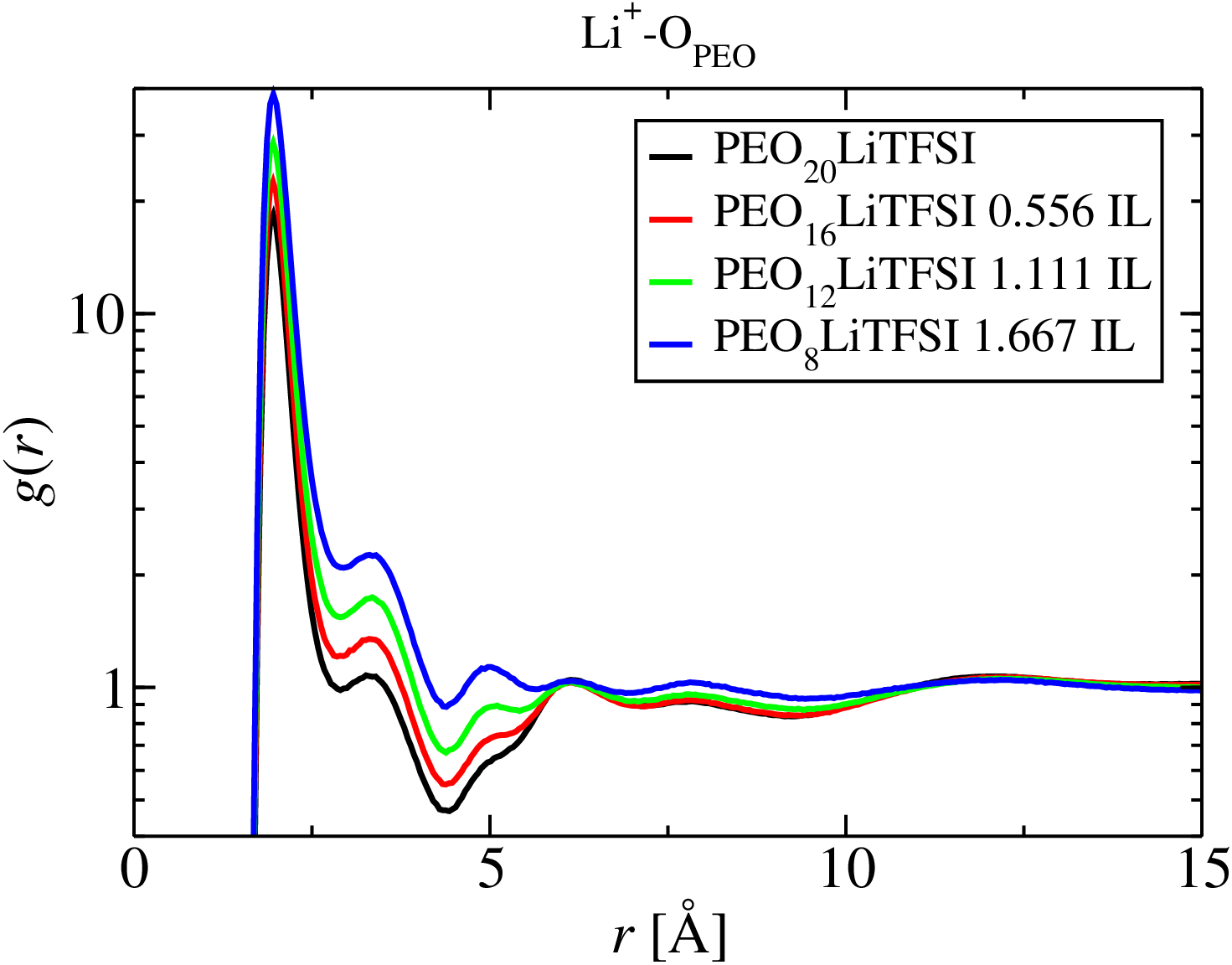} }
 \subfigure[ \label{fig:gofr_LiO_TFSI_xminus} ]{ \includegraphics[scale=0.3]{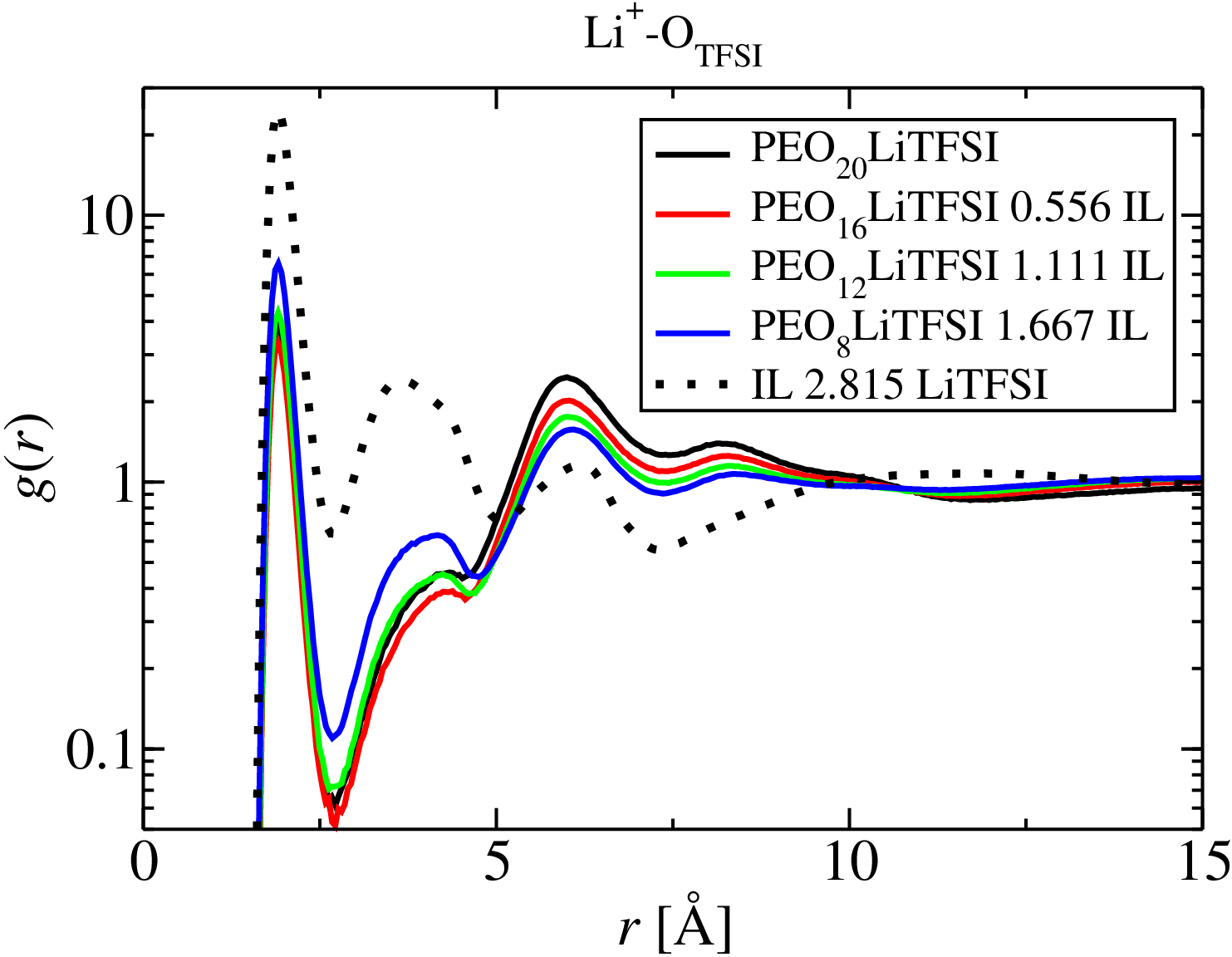} }
 \caption{Upper: radial distribution functions in \Px\ for the atom pairs (a)~\LiOPEO\ and (b)~\LiOTFSI. 
          Lower: radial distribution functions in \Pminusx\ for the atom pairs (c)~\LiOPEO\ and (d)~\LiOTFSI. }
 \label{fig:gofr_Li}
\end{figure*}

Figure~\ref{fig:gofr_Li} shows the radial distribution functions $g(r)$ for the atom pairs \Li$-\text{O}_\text{PEO}$ and \Li$-\text{O}_\text{TFSI}$. 
Both coordination types exhibit a sharp peak around $2\text{~\AA}$ corresponding to the first coordination shell. 
For \LiOPEO, the first coordination sphere directly crosses over into a second small peak around $3.3\text{~\AA}$. 
At larger distances no significant structural arrangement can be found. 
For the coordination of \LiOTFSI, peaks become noticeable also at larger distances around $6\text{~\AA}$ and $8\text{~\AA}$, thus demonstrating 
that long-ranged correlations are present as also observed for a PEO$_{20}$LiI electrolyte~\cite{MaitraJPCB2008}. 
The same observation can be made for the PEO-free electrolyte \Pzero, as also reported in previous MD studies~\cite{BorodinJPCB2006_ILLiTFSI}. 
The pair correlation function of the PYR$_\mathrm{13}$ cations and TFSI anions (\ie \MPPYOTFSI) exhibits only a weak first coordination peak 
between $3.6$ and $5.0\text{~\AA}$. 

When successively adding the IL, one observes that the peak positions of both \LiOPEO\ and \LiOTFSI\ remain the same for all electrolytes. 
For \Px, the EO coordination numbers extracted from the integral over the first shell increase slightly, partly as a result of the lower 
fraction of lithium ions coordinating to two PEO chains (Table~\ref{tab:structure} in the main article), for which the EO coordination number is 
lower due to steric effects. 
In \Pminusx, the absolute EO coordination number (as determined from the integral over $g(r)$) decreases with increasing IL content. 
This trend can be explained by the decreasing number of possible coordination sites, since the PEO concentration reduces. 
Naturally, the coordination number of the TFSI oxygens increases with increasing IL content in both types of electrolytes.

\subsection{Lithium Coordination Shell}

\begin{figure*}
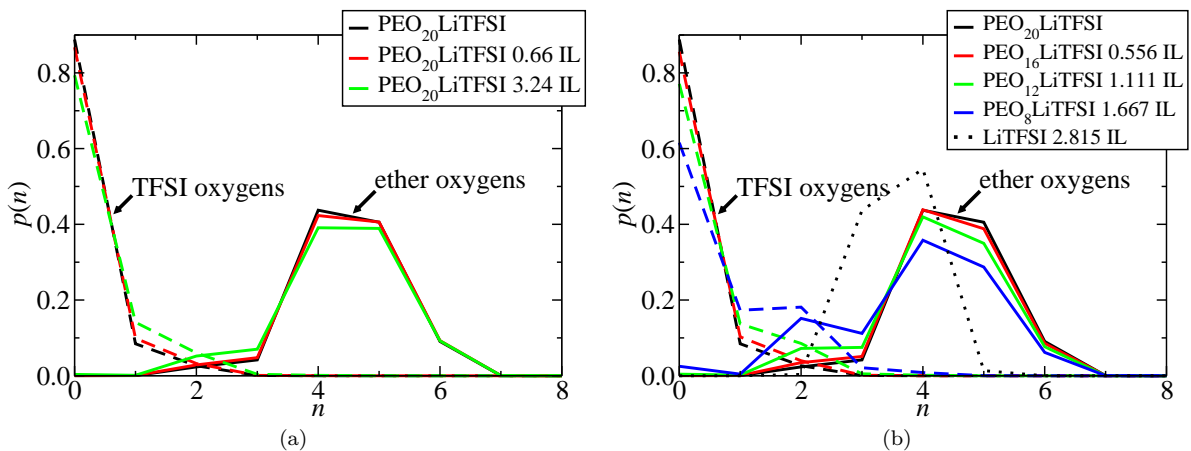

 \centering
 \subfigure[ \label{fig:EO_bin_x} ]{ \includegraphics[scale=0.3]{PILbin_add} }
 \subfigure[ \label{fig:EO_bin_xminus} ]{ \includegraphics[scale=0.3]{PILbin_sub} }
 \caption{Probability $p(n)$ to find a certain coordination number $n$ of EOs 
          (irrespective if the ion is tied to one or two PEO chains) or TFSI oxygens
          in (a)~\Px\ and (b)~\Pminusx. }
 \label{fig:coordbins}
\end{figure*}

Figure~\ref{fig:coordbins} shows the probability distribution functions $p(n)$ to find a lithium ion with $n$ EOs or TFSI oxygens in its first coordination shell. 
One observes that for \PxIL, the coordination numbers are very similar in all systems and only change slightly with the IL concentration. 
In contrast to this, for \Pminusx, a coordination number of $n=2$ EOs becomes noticeable, and simultaneously the probability for one or two coordinating 
anions increases.

\section{Part~B: Force Field Validation}
\label{sec:validation}

In order to validate the results from the AMBER simulations, comparative simulations with Lucretius~\cite{Lucretius} (the original code for which the force 
field used in this study has been developed) have been performed. 
We focus on the pure polymer electrolyte \Ptwenty. 
For the Lucretius simulation, the same techniques and parameters as in the original study~\cite{BorodinMamol2006} have been used. 
Both MD codes yielded the same density when simulating an $NpT$ ensemble, thus giving a first indication 
that the AMBER results are reasonable. 
In the following, $NVT$ simulations of the respective codes will be compared 
(the results are the same for $NpT$ simulations). 
Since the Lucretius version used in this comparison (\ie the same as in ref.~\citenum{BorodinMamol2006}) does not allow parallel calculations, the total 
simulation length was restricted to $50\text{~ns}$ as compared to the $200\text{~ns}$ of the AMBER simulations (note that in the meantime a newer, parallelized 
Lucretius version is also available).

\subsection{Dihedral Distributions}

\begin{figure*}
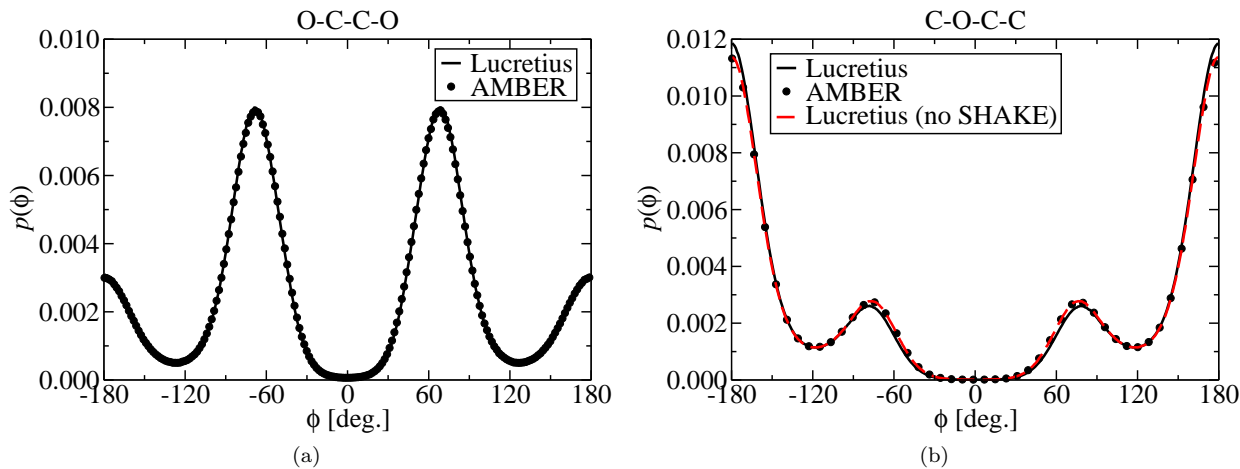

 \centering
 \subfigure[ \label{fig:dihedral_OCCO} ]{ \includegraphics[scale=0.3]{dihedral_O-C-C-O_bin_val} }
 \subfigure[ \label{fig:dihedral_COCC} ]{ \includegraphics[scale=0.3]{dihedral_C-O-C-C_bin_val} }
 \caption{Probability distribution function of two dihedral types of the PEO backbone for \Ptwenty\ at 
          $T=423\text{~K}$. (a) O-C-C-O and (b) C-O-C-C. }
 \label{fig:dihedral_bin}
\end{figure*}

Figure~\ref{fig:dihedral_bin} shows the probability distribution function $p(\phi)$ of the dihedral angles 
$\phi$ describing the rotation around the carbon-carbon ($\text{O-C-C-O}$) and the carbon-oxygen ($\text{C-O-C-C}$) 
bonds of the PEO backbone. 
For the $\text{O-C-C-O}$-dihedral, one observes perfect agreement between the two codes, whereas for the conformations 
of the $\text{C-O-C-C}$-dihedral slight differences can be observed. 
These deviations can be rationalized by the fact that the bond length constraining (SHAKE~\cite{SHAKE}) for non-hydrogen 
bonds is not possible in parallel AMBER simulations. 
Since these bonds can be stretched in these simulations, the $1$-$4$ repulsions between the outer carbon atoms of the 
$\text{C-O-C-C}$-dihedral is mitigated, and the {\it gauche}-conformation ($\phi\approx 75^\circ$) becomes more populated 
compared to the {\it trans}-conformation ($\phi=180^\circ$). 
Indeed, the same observation can be made for a Lucretius simulation (length $25\text{~ns}$) without bond length constraining. 
However, these slight deviations have no influence on the dynamics of the PEO backbone and the lithium motion (see discussion below).

\subsection{Radial Distribution Functions}

\begin{figure*}
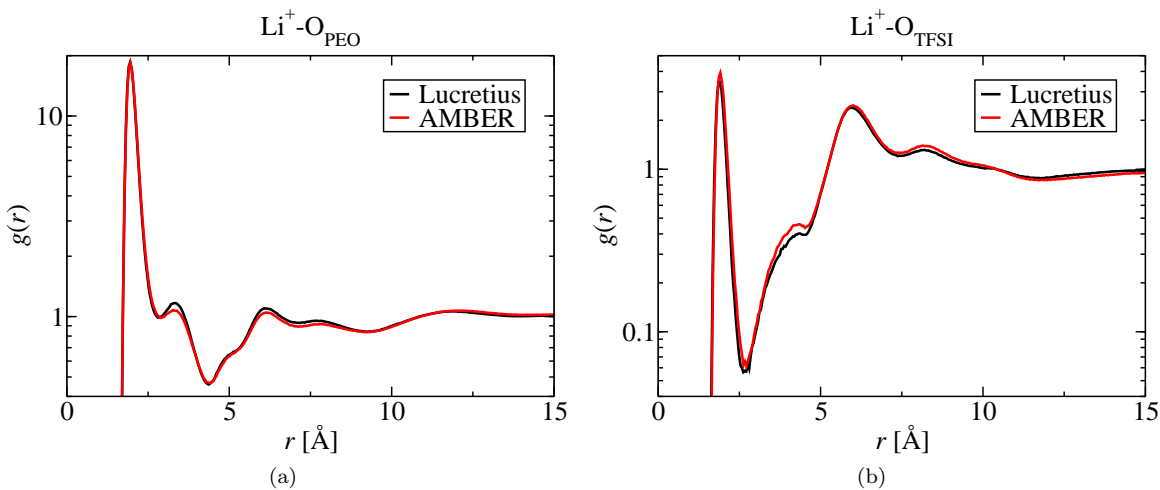

 \centering
 \subfigure[ \label{fig:gofr_validation_PEO_Li} ]{ \includegraphics[scale=0.3]{gofr_PEO-O_Li_val} }
 \subfigure[ \label{fig:gofr_validation_TFSI_Li} ]{ \includegraphics[scale=0.3]{gofr_TFSI-O_Li_val} }
 \caption{Radial distribution functions $g(r)$ of the 
          (a)~\Li$-\text{O}_\text{PEO}$ and the 
          (b)~\Li$-\text{O}_\text{TFSI}$ interaction. }
 \label{fig:gofr_validation}
\end{figure*}

Figure~\ref{fig:gofr_validation} shows the radial distribution functions $g(r)$ for the \Li$-\text{O}_\text{PEO}$ and the 
\Li$-\text{O}_\text{TFSI}$ interaction. For both coordination types one observes reasonable agreement. 
Again, the slight deviations for the second coordination peak are related to the fact that the molecules in AMBER are more 
flexible due to the unconstrained bonds.

\subsection{Mean Square Displacements}

\begin{figure}
 \centering
 \includegraphics[scale=0.3]{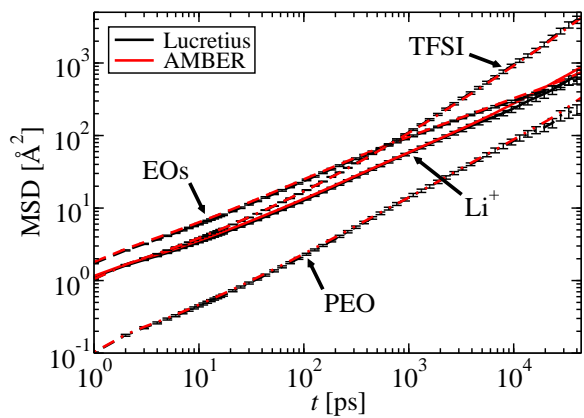}
 \caption{MSDs for the individual molecule types in \Ptwenty\ simulated with Lucretius and AMBER. }
 \label{fig:msd_validation}
\end{figure}

Figure~\ref{fig:msd_validation} shows the MSDs for the individual molecular species in \Ptwenty. 
Since the AMBER simulation extends over $200\text{~ns}$ and the MSDs thus have much smaller error bars, only the error 
bars of the Lucretius curves are shown for clarity. 
For PEO, an additional distinction was made between the segmental dynamics (EOs) and center-of-mass motion. 
The agreement for all molecule types is quantitative. 
Deviations for larger time scale of several ten nanoseconds are within the error bars of the shorter Lucretius simulation. 
In total, the agreement between the two MD codes is convincing.

\section{Part~C: Benchmark Calculations}
\label{sec:benchmark}

Benchmark calculations have been performed for the modified AMBER code. 
Again, we focus on the electrolyte \Ptwenty\ ($4772$ atoms) at $T=423\text{~K}$ in the $NVT$ ensemble 
using various numbers of CPUs (Intel Xeon Nehalem X5550 with $2.67\text{~GHz}$ each, PALMA cluster~\cite{PALMA}). 
The simulations were carried out using various treatments of the polarization, in particular without polarization, iterative solution 
of the inducible dipoles, as well as the Car-Parrinello scheme described in ref.~\citenum{vanBelleMolPhys1992}. 

\begin{figure*}
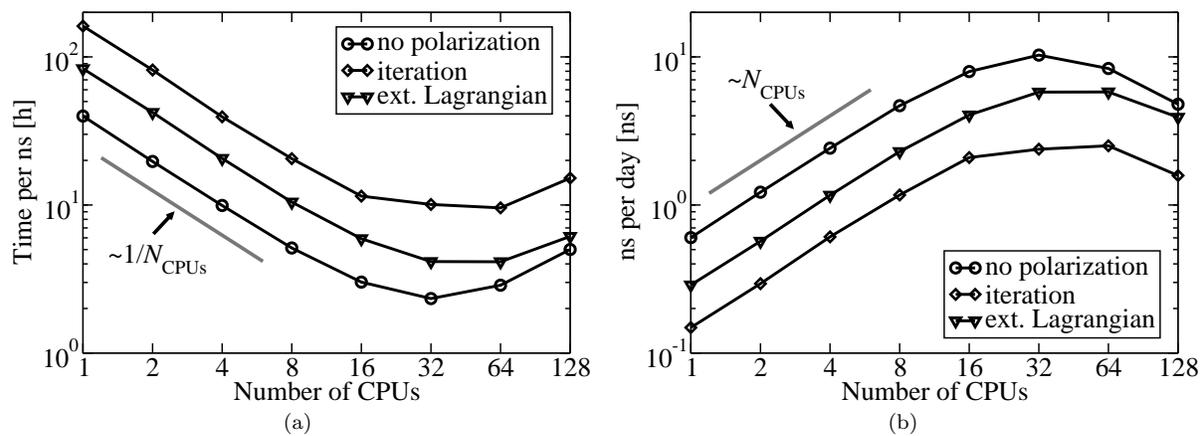

 \centering
 \subfigure[ \label{fig:hour_per_ns} ]{ \includegraphics[scale=0.3]{time_per_ns_hours} }
 \subfigure[ \label{fig:ns_per_day} ]{ \includegraphics[scale=0.3]{nanoseconds_per_day} }
 \caption{Computation time for \Ptwenty\ ($4772$ atoms) in dependence of the number of CPUs for various polarization treatments. 
          For convenience, the accessible MD time per day is also shown. }
 \label{fig:benchmark}
\end{figure*}

The computation time is shown in Figure~\ref{fig:benchmark}. 
For convenience, the resulting MD time per day is also shown in Figure~\ref{fig:benchmark}. 
For small processor numbers, one observes an ideal linear increase of the computation speed. 
From about eight CPUs this scaling drops below linear scaling, and finally saturates at a maximum computation speed for $32\text{~CPUs}$. 
A more detailed analysis of the computation times of the distinct contributions (\ie bonded, van-der-Waals or electrostatic interactions) 
reveals that the saturation as well as the subsequent decrease in Figure~\ref{fig:benchmark} arises from the evaluation of the reciprocal 
part of the electrostatic energies and forces. 
For this step, all spatial coordinates of the system are involved in the calculation, requiring the excessive communication between the 
individual CPUs. 
Consequently, the speed-up gained by the larger number of CPUs cannot compensate for this massive broadcast, so that the real computation 
time saturates and finally decreases again. 

The incorporation of polarization leads to a decrease of the simulation speed. 
Here, the Car-Parrinello-like treatment of the dipoles~\cite{vanBelleMolPhys1992} is significantly faster than the iterative solution of 
the inducible dipoles. 
For many CPUs, their relative difference becomes even larger. 
As discussed above, this effect is related to the enormous broadcast and the exchange of the inducible dipoles between the individual CPUs. 
This slowdown becomes especially heavy for the iterative scheme, as the dipoles have to be gathered and redistributed on the nodes in each 
iteration step. 
\end{appendix}

\bibliography{thebib}

\end{document}